\documentclass[useAMS,usenatbib,usegraphicx]{mn2e}

\usepackage{paper_def}
\usepackage{txfonts}
\usepackage[T1]{fontenc}
\usepackage{aecompl}
\usepackage{color}
\usepackage{natbib}
\usepackage{appendix}
\usepackage{threeparttable}
\usepackage{longtable}
\usepackage{pifont}
\newcommand{\cmark}{\ding{51}}%

\newcommand{\lsun}{L$_{\odot}$}
\newcommand{\sfr}{M$_{\odot}$ yr$^{-1}$}

\newcommand{\lfir}{${\rm L_{FIR} }$}

\newcommand{\cii}{[C{\sc ii}]}

\newcommand{\lcii}{L$_{\rm[CII]}$}
\newcommand{\ecii}{$\Sigma_{\rm[CII]}$}
\newcommand{\esfr}{$\Sigma_{\rm SFR}$}
\newcommand{\lya}{Ly$\alpha$}
\newcommand{\Td}{T$_{\rm d}$}

\title[Kiloparsec-scale gaseous clumps and star formation at $z$=5--7]{Kiloparsec-scale gaseous clumps and star formation at $z$=5--7}
\author[Carniani et al.]{S.~Carniani,$^{1,2}$
R.~Maiolino,$^{1,2}$ 
R. Amor\'in,$^{1,2}$ 
L.~Pentericci,$^3$ 
A.~Pallottini,$^{1,2,4,5}$
\newauthor
A.~Ferrara,$^5$
C.~J.~Willott,$^{6}$
R.~Smit,$^{1,2}$
J.~Matthee,$^{7}$
D.~Sobral,$^{7,8}$
 P.~Santini,$^3$
 \newauthor
 M.~Castellano,$^3$
  S.~De~Barros,$^{9,10}$
 A.~Fontana,$^3$
 A.~Grazian,$^3$
 and L.~Guaita$^{11}$
\\
$^{1}$Cavendish Laboratory, University of Cambridge, 19 J. J. Thomson Ave., Cambridge CB3 0HE, UK\\
$^{2}$Kavli Institute for Cosmology, University of Cambridge, Madingley Road, Cambridge CB3 0HA, UK\\
$^{3}$INAF - Osservatorio Astronomico di Roma, via Frascati 33, 00078 Monteporzio, Italy\\
$^{4}$Centro Fermi, Museo Storico della Fisica e Centro Studi e Ricerche `Enrico Fermi', Piazza del Viminale 1, Roma, 00184, Italy \\
$^{5}$Scuola Normale Superiore, Piazza dei Cavalieri 7, 56126 Pisa, Italy\\
$^{6}$ NRC Herzberg, 5071 West Saanich Rd, Victoria, BC V9E 2E7, Canada\\
$^{7}$Leiden Observatory, Leiden University, PO Box 9513, NL-2300 RA, Leiden, The Netherlands\\
$^{8}$Department of Physics, Lancaster University, Lancaster, LA1 4YB, UK \\
$^{9}$Observatoire de Gen\`eve, Universit\'e de Gen\`eve, 51 Ch. des Maillettes, 1290 Versoix, Switzerland\\
$^{10}$ INAF--Osservatorio Astronomico di Bologna, via P. Gobetti 93/3, I-40129,
Bologna, Italy \\
$^{11}$ N\'ucleo de Astronom\'ia, Facultad de Ingenier\'ia, Universidad Diego Portales, Av. Ej\'ercito 441, Santiago, Chile
}

\begin{document}


\pagerange{\pageref{firstpage}--\pageref{lastpage}} \pubyear{2002}

\maketitle

\label{firstpage}

\begin{abstract}
We investigate the morphology of the \cii\ emission in a sample of ``normal''
star-forming galaxies at $5<z<7.2$
in relation to their UV (rest-frame) counterpart.
We use new ALMA observations of galaxies at $z\sim 6-7$, as well as a careful
re-analysis of archival ALMA data.
In total 29 galaxies were analysed, 21 of which are detected in \cii.
For several of the latter the \cii\ emission breaks into multiple components.
Only a fraction of these \cii\ components, if any, is associated with the primary UV systems,
while the bulk of the \cii\ emission is associated either with fainter UV components, or not associated with any UV counterpart at the current limits.
By taking into account the presence of all these components,
we find that the \lcii-SFR relation at early epochs is fully consistent
with the local relation, but it has a dispersion of 0.48$\pm$0.07 dex, which
is about two times larger than observed locally.
We also find that the deviation from the local \lcii-SFR relation has a weak anti-correlation
with the EW(\lya).
The morphological analysis also reveals that  \cii\ emission is generally much more extended than the UV emission. As a consequence,
these primordial galaxies are characterised by a \cii\ surface brightness generally much lower than
expected from the local $\rm \Sigma _{[CII]}-\Sigma _{SFR}$ relation.
These properties are likely a consequence of a combination of different
effects, namely: gas metallicity, \cii\ emission from
 obscured star-forming regions, strong variations of the ionisation parameter, and
circumgalactic gas in accretion or ejected by these primeval galaxies.
\end{abstract}


\section{Introduction}

The morphological  investigation of  galaxies in the early Universe can provide important information on their
formation and evolutionary processes.
For instance, a disturbed and multi-clump morphology, especially in the cold phase, may suggest
the presence of  disc instabilities, and can give indication of feedback processes, as well as minor and/or major merger events during the galaxy assembly \citep[e.g.,][]{Tamburello:2015, Fiacconi:2016, Ceverino:2017, Pallottini:2017,Pallottini:2017a}.
In this context, the advent of facilities delivering high angular resolution observations has enabled us to probe the internal structure of galaxies  in the distant Universe, revealing that clumpy morphologies are more common at higher redshift than at $z=0$ \citep[e.g.][]{Forster-Schreiber:2006,Genzel:2008}. 
For instance, the fraction of  galaxies at $0.5<z<3$  exhibiting kpc-scale clumps with sustained star-formation activity is higher than 30\% \citep{Ravindranath:2006, Elmegreen:2009, Guo:2015}.
At higher redshifts the fraction of galaxies showing disturbed morphology or multi clumps is even higher. on On a sample of 51 Ly$\alpha$ emitters (LAEs) and 16 Lyman break galaxies (LBGs) at $z>5.7$ \cite{Jiang:2013} found that roughly half of the brightest galaxies ($M_{1500}$<-20.5 mag) are made of multiple components that may be merging.
Near-infrared (NIR) high-resolution imaging have also revealed that irregular shapes with multi-clump morphology are prevalent in LAEs and LBGs at $z\sim7$ within the epoch of reionization \citep{Ouchi:2010, Sobral:2015, Matthee:2017, Bowler:2017}.

Additional identification of clumpy systems in the early epoch and a detailed characterisation of  high-$z$ clumps or satellites is fundamental to constrain galaxy assembly. 
The extended Atacama Large Millimetre/submillimetre Array (ALMA) configurations enable us to reach high-angular
resolution
and exploit far-infrared  (FIR) fine structure emission lines, such as \cii\ at 158$\mu$m, as
powerful diagnostics to assess the morphology of primeval galaxies. \cii\  is emitted primarily in the (mostly neutral) atomic and molecular gas associated with Photon
Dominated Regions (excited by the soft UV photons), but also in partly ionised regions and it is one of the
primary coolants of the ISM. Indeed, it is generally the strongest emission lines observed in the spectra of
galaxies. Since its first detection at high redshift \citep{Maiolino:2005} this transition has then been
detected in large samples of distant galaxies. However, until recently, the \cii\ emission was only detected
in extreme environments, such as quasar host galaxies and SMGs, characterised by SFRs of several hundred
solar masses per year, not really representative of the bulk of the galaxy population at these epochs
\citep[e.g.][]{Maiolino:2009, Maiolino:2012, De-Breuck:2011, Wagg:2012, Gallerani:2012, Carilli:2013, Carniani:2013, Williams:2014, Riechers:2014, Yun:2015, Schreiber:2017, Trakhtenbrot:2017, Decarli:2017}.
Detecting \cii \ in ``normal'' galaxies has required the sensitivity delivered by ALMA.
To date, the \cii\ line has been detected in several galaxies at $z>5$  and it is  spatially resolved in most of these targets \citep{Capak:2015,Willott:2015, Maiolino:2015, Knudsen:2016, Pentericci:2016, Bradac:2016, Smit:2017, Carniani:2017a,Matthee:2017}. 
\cite{Smit:2017} recently presented \cii\ observations of two galaxies at $z\sim7$ characterised by a gradient of velocity consistent with a undisturbed  rotating gas disk.
However, most of the $z>5$ galaxies show  extended and clumpy \cii\ emission  with velocities consistent with the systematic redshift of the galaxy ($|\Delta$v$_{\rm Ly\alpha}|<500$~km/s) but spatially offset relative to the rest-frame UV counterpart \citep{Maiolino:2015, Willott:2015, Capak:2015,Carniani:2017a, Carniani:2017, Jones:2017}.
In many cases these offsets have been ignored or ascribed to astrometric uncertainties. However, based
on detailed astrometric analysis,
it has been shown that most of these offsets are physical (a revised analysis will be given in this paper),
hence they should be taken as an important signature of the evolutionary processes in the early phases of
galaxy formation.
%
Various scenarios have been proposed to explain the positional offsets between \cii\ and star-forming regions such as
stellar feedback clearing part of the  ISM, gas accretion, wet mergers, dust obscuration and variations of the
ionisation parameter \citep[e.g.][]{Vallini:2015, Katz:2017}. 
\cite{Barisic:2017} and \cite{Faisst:2017a} have recently found (rest-frame) UV faint companions  whose locations is
consistent with the displaced \cii\ emission, suggesting that the carbon line traces star-forming regions where the UV light is absorbed by dust.

As mentioned,
most previous studies have attempted to assess the nature of \cii\ emission  in primeval galaxies neglecting the positional offsets between the FIR line and rest-frame UV emission.  
The goal of this paper is to assess the connection between \cii\ and SFR in the early Universe by taking into
account the multi-clump morphology of galaxies at $z>5$ and by associating the components with their
proper optical-UV counterparts (if detected). This is achieved by re-analysing ALMA \cii\ observations of $z > 5$ star-forming galaxies, and by performing a detailed kinematical analysis of the \cii\ line, in order to deblend the different components of the multi-clump systems. 
In addition to previous ALMA observations, partly discussed in literature, we also make use of new ALMA data
targeting five $z\sim6$ star-forming galaxies with SFR~$<20$~\sfr. In Section~\ref{sec:sample} we detail the  sample
and the analysis of ALMA observations. The morphological analysis is presented in Section~\ref{sec:Multi-component
systems}, while the relation between the \cii\ and SFR is discussed in Section~\ref{sec:dispersion}. In
Section~\ref{sec:lyaEW} we investigate the connection between  \cii\ luminosity and \lya\ strength in our sample.
Section~\ref{sec:spatially}  focuses on the spatial extension of the \cii\ and UV emission
and the correlation between \cii\ surface brightness and SFR surface density. We discuss the findings in Section~\ref{sec:discussion}, while the conclusions of this work are reported in  Section~\ref{sec:conclusions}

Throughout this paper we assume the following cosmological parameters: $H_0 = 67.8$ km s$^{-1}$ Mpc$^{-1}$ , $\Omega_M = 0.308$, $\Omega_\Lambda$ = 0.685 \citep{Planck:2016}

\section{Sample, observations and analysis}\label{sec:sample}

\subsection{Archival data}\label{sec:literature_sample}

%
%
%
\begin{table}
 \centering
  \caption{Overview of the $z>5$ star-forming galaxies observed with ALMA used in this paper, ordered by name.}
 
 \addtolength{\tabcolsep}{-3pt}    
  \begin{tabular}{lccccc }
  \hline
  \hline
   Target$^{(a)}$ & Ra$^{(b)}$ & Dec$^{(c)}$ &   Ref.$^{(d)}$  & \cii$^{(e)}$ & Clumpy$^{(f)}$  \\
  \hline
 	\multicolumn{6}{c}{\emph{Literature Sample}} \\
	  \hline
BDF3299 & 337.0511  & -35.1665 & 1,2 &  \cmark & \cmark\\
BDF512 & 336.9444 & -35.1188 & 1 & &  \\
CLM1 & 37.0124 &  -4.2717 & 3 &  \cmark &  \\
COSMOS13679 & 150.0990 & 2.3436 & 4 & \cmark & \\
COSMOS24108 & 150.1972 & 2.4786 & 4 & \cmark & \cmark\\
COS-2987030247 & 150.1245 & 2.2173 & 5 & \cmark &  \\
COS-3018555981 & 150.1245 & 2.2666 & 5 & \cmark &  \\
CR7   & 150.2417 & 1.8042 & 6 &  \cmark & \cmark\\
Himiko & 34.4898 &  -5.1458 & 7,12 & \cmark & \cmark\\
HZ8 & 150.0168 & 2.6266 & 8 & \cmark  & \cmark \\
HZ7 & 149.8769 & 2.1341 &  8 & \cmark  &   \\
HZ6 & 150.0896 & 2.5864 & 8 & \cmark & \cmark \\
HZ4 & 149.6188 & 2.0518 & 8 & \cmark &  \\
HZ3 & 150.0392 & 2.3371 & 8 & \cmark &  \\
HZ9 & 149.9654 & 2.3783 & 8 & \cmark &  \\
HZ10 & 150.2470 & 1.5554 & 8  & \cmark & \cmark \\
HZ2 & 150.5170 & 1.9289 & 8 & \cmark & \cmark \\
HZ1 & 149.9718 & 2.1181 & 8  & \cmark &  \\
IOK-1 & 200.9492 & 27.4155 & 9 & &  \\
NTTDF6345 & 181.4039 & -7.7561 & 4 & \cmark &  \\
SDF46975 & 200.9292 & 27.3414  &  1 &  \\
SXDF-NB1006-2 & 34.7357  & -5.3330 & 10 & & \\
UDS16291 & 34.3561 & -5.1856 & 4 &  \cmark & \\
WMH5 &  36.6126 & -4.8773 & 3,11 &  \cmark & \cmark\\
      \hline
 	\multicolumn{6}{|c|}{\emph{Additional new data}} \\
	  \hline
BDF2203 & 336.958 & -35.1472 &  & \cmark &  \\
GOODS3203 & 53.0928 & -27.8826 \\
COSMOS20521  & 150.1396 & 2.4269 &  & &  \\
NTTDF2313 & 181.3804 & -7.6935 &  &  \\
UDS4812 & 34.4768 & -5.2472 &  &  & \\

 \hline
\end{tabular}
\\

\begin{tablenotes}[flushleft]
\footnotesize
\item  {\bf Notes}.
{\bf (a)} Name of the source.
{\bf(b, c)} J2000 coordinates.
{\bf(d)} References in which  ALMA observations are presented (
[1] \citealt {Maiolino:2015}; [2] \citealt{Carniani:2017a};
[3] \citealt{Willott:2015}; [4] \citealt{Pentericci:2016};
[5] \citealt{Smit:2017}; [6] \citealt{Matthee:2017};
[7] \citealt{Ouchi:2013}; [8] \citealt{Capak:2015};
[9] \citealt{Ota:2014}; [10] \citealt{Ouchi:2013};
[11] \citealt{Jones:2017} [12] Carniani et al. 2017b)
{\bf(e)} Check mark, \cmark, indicates that \cii\ emission has been detected at the redshift of the galaxy.
{\bf(f)} Check mark, \cmark, indicates that the galaxy has a clumpy morphology.

\end{tablenotes}
\label{tab:projects}
\end{table}

The sample is mainly drawn from the archive and literature by selecting only spectroscopically confirmed
star-forming galaxies at $z>5$ observed with ALMA in the \cii\ line. 
We limit our sample to those systems with  SFR  $\lesssim
100$ \sfr\ since they are representative of the bulk of the galaxy population in the  early Universe \citep[e.g.][]{Robertson:2015, Carniani:2015}.
The list of selected sources is given in Table~\ref{tab:projects}.
The  sample does not include lensed systems \citep{Knudsen:2016, Gonzalez-Lopez:2014, Bradac:2016, Schaerer:2015}
since  magnification factor uncertainties may lead to large errors on SFR and \cii\ luminosity estimates, as well
as on the morphology analysis.

For the purpose of our investigation, which focuses on the nature and implications of the positional offsets
between \cii\ and UV emission, we have retrieved and re-analysed  ALMA data revealing a \cii\ detection at the systemic velocity of the galaxy (see Table~\ref{tab:projects}). For these objects,  ALMA observations have been calibrated following the prescriptions presented in  previous  works. 

In addition to the rest-frame FIR images, we have also used Hubble Space Telescope ({\it HST}) and Visible and Infrared Survey Telescope for Astronomy (VISTA) NIR observations (rest-frame UV at $z>5$). 
ALMA, {\it HST} and VISTA data have been aligned based on the location of serendipitous sources detected in both
ALMA continuum and NIR  images by  assuming that the  millimetre emission of these sources is cospatial to
the near-infrared map. This is also supported by the fact that all foreground sources used for registering
millimetre and NIR images do not exhibit any   multi-clump or merger-like morphologies indicating that
astrometric offsets between the ALMA and NIR images are likely associated to astrometric calibrations. 
For those observations revealing the presence of two (or more) serendipitous sources  we have verified that the
astrometric shift for each source is consistent with that estimated from the other source(s) in the same map.
In all cases we have checked that the estimated astrometric offset is consistent with those obtained by aligning NIR foreground sources and ALMA phase calibrators to their astrometric position from the GAIA Data Release 1 catalogue \citep{Gaia-Collaboration:2016}. %
For those sources whose ALMA continuum map showing no serendipitous sources, we have matched the NIR foreground sources and ALMA phase calibrators to either GAIA Data Release~1 catalogue \citep{Gaia-Collaboration:2016} or  AllWISE catalogue \citep{Cutri:2013}. 
We note that in all ALMA datasets the locations of the various phase calibrators are in agreement within the
error with the GAIA  and  AllWISE catalogues, implying that, when a systemic (not physical)
offset is seen, this is generally due to some small astrometric uncertainties in the optical-NIR
data.
These astrometric issues have  been also discussed by \cite{Dunlop:2016} who analysed ALMA images targeting the Hubble Ultra Deep Field.
We therefore applied the  astrometric shifts, which span a range between 0.1\arcsec\ and 0.25\arcsec,  to the NIR images.

We note that additional systems with low SFRs have also been tentatively
detected in 14 \cii\ line emitting candidates at $6 < z < 8$ \citep{Aravena:2016a},
which are not included in this analysis as they are not spectroscopically confirmed
yet and $\sim$60\% of these objects are expected to be spurious .
%


\subsection{Additional new ALMA data}
\label{sec:additional}

In addition to the archival/literature sample, we  have also included new  \cii\ observations of five star-forming
galaxies  at $z\sim6$  with a SFR$\sim$10 \sfr\  observed with ALMA in Cycles 3 and 4 (P.I.
Pentericci). The five new sources, listed in Table~\ref{tab:projects}, have been selected from a sample of
$>120$ LBGs at $z\sim6-7$. These have been spectroscopically confirmed thanks to  recent ultra-deep spectroscopic
observations from the ESO Large Program CANDELSz7  \citep[P.I. L. Pentericci;][]{DE-Barros:2017}. The proposed
ALMA programs aimed at observing \cii\ emission in ten  star-forming galaxies at $z>6$ with UV luminosities
lower than -21 mag (UV SFR $<20$ \sfr) and spectroscopic redshift uncertainties $<$0.03, but only five
galaxies have been observed. The observations and data calibrations are presented in
the Appendix~\ref{sec:appA}. We have registered NIR images to ALMA observations by matching the location of
the foreground, serendipitous continuum
sources  and ALMA calibrators to the position given by the GAIA Data Release 1 catalogue (Gaia Collaboration et al. 2016).

While the continuum emission is not detected at the location of any of the five galaxies, we detect two  and one serendipitous sources in the COSMOS20521 and  BDF2313 continuum maps, respectively. 
The positions of the serendipitous continuum sources is in agreement with the location of NIR foreground galaxies, thus confirming the astrometric shifts estimated from the catalogue.

\begin{figure}
\centering
\includegraphics[width=0.8\columnwidth]{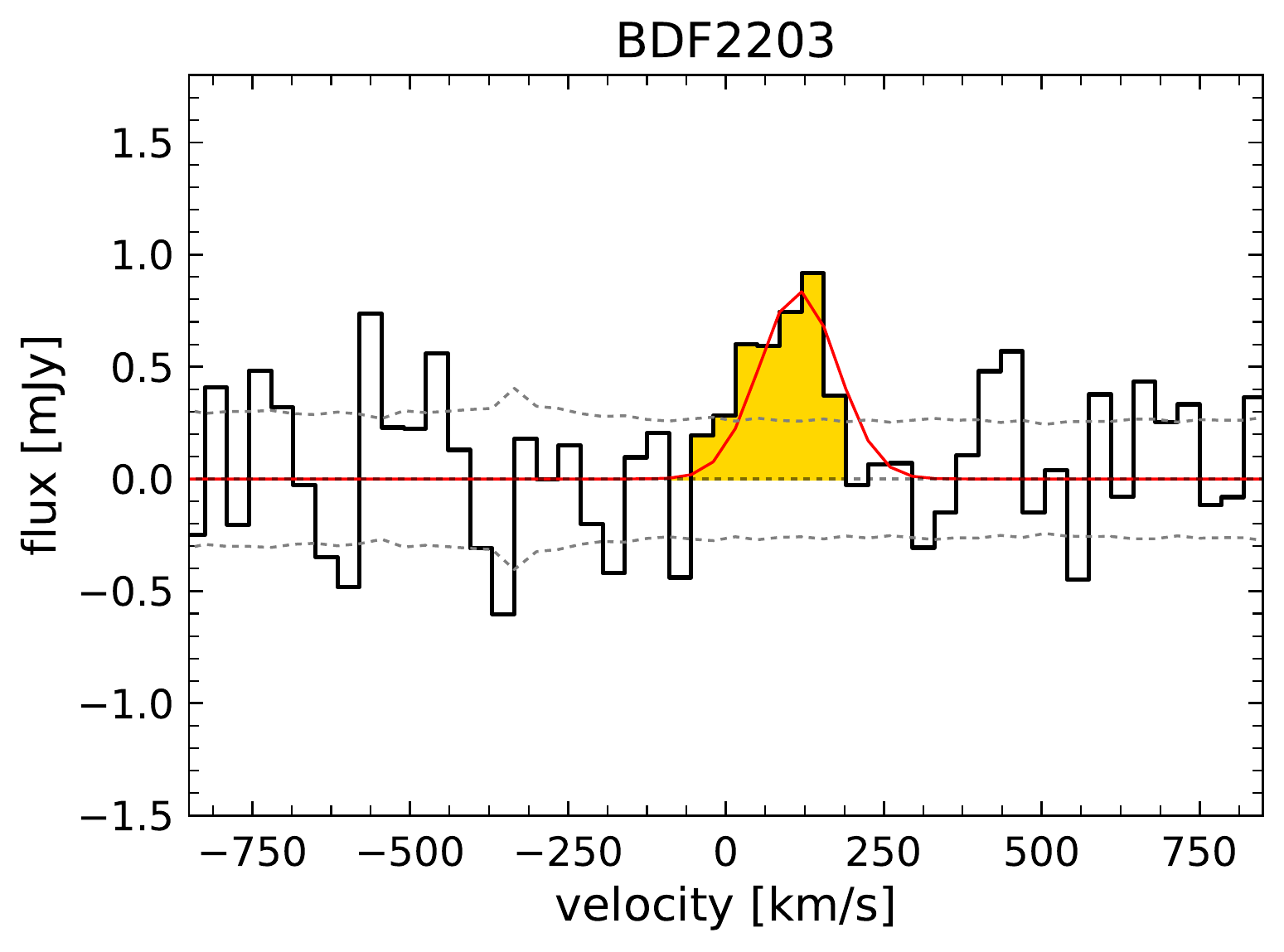}  %
\caption{ ALMA spectrum of BDF2203 showing a new \cii\ detection. The velocity reference is set to the redshift defined by the Ly$\alpha$.
The dotted grey lines shows the 1$\sigma$ and -1$\sigma$ per channel. The red line  indicates the best fit 1D Gaussian line profile.} 
 \label{fig:appFig1}
\end{figure}

\begin{figure}
\centering
\includegraphics[width=0.6\columnwidth]{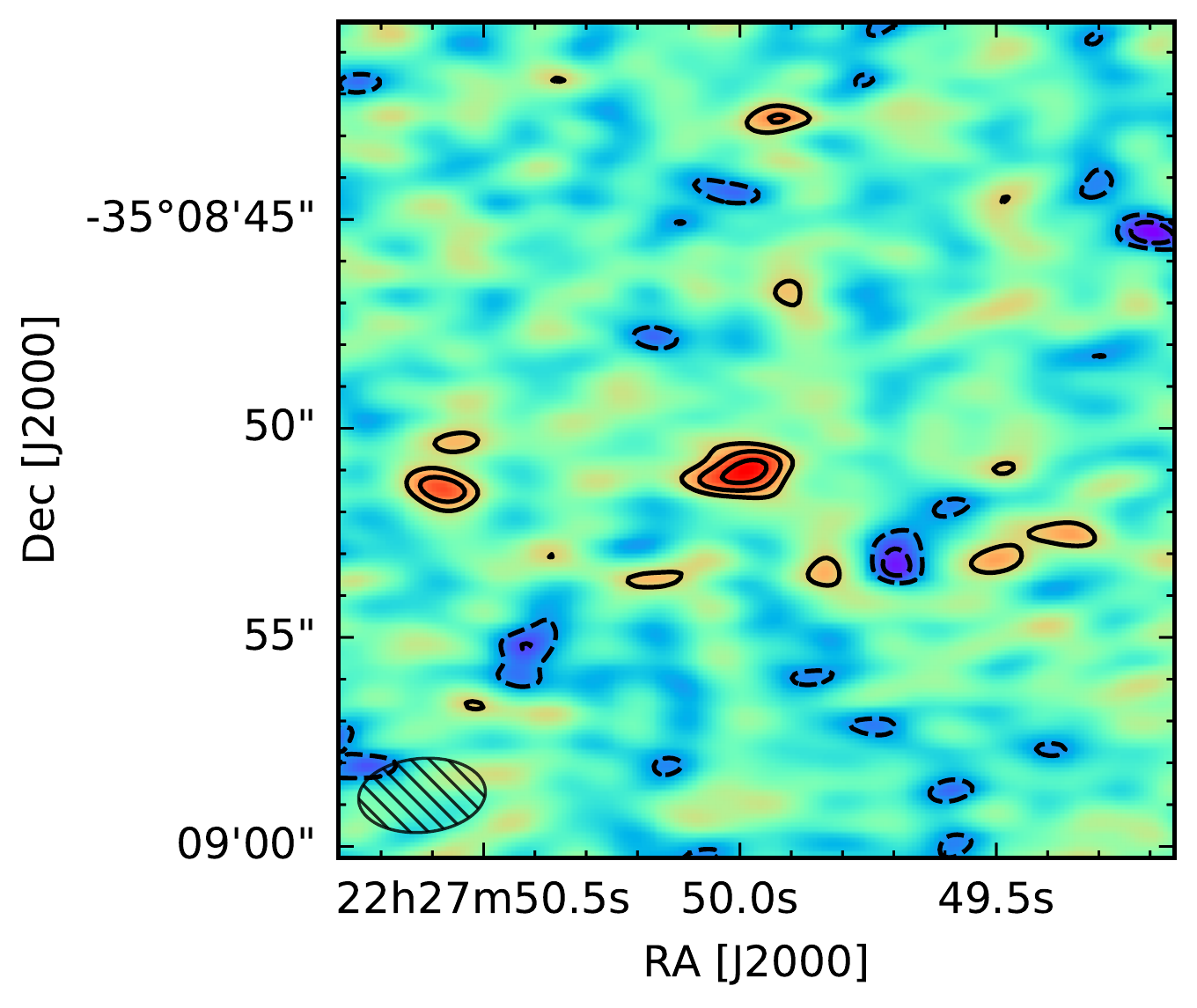}  %
\includegraphics[width=0.6\columnwidth]{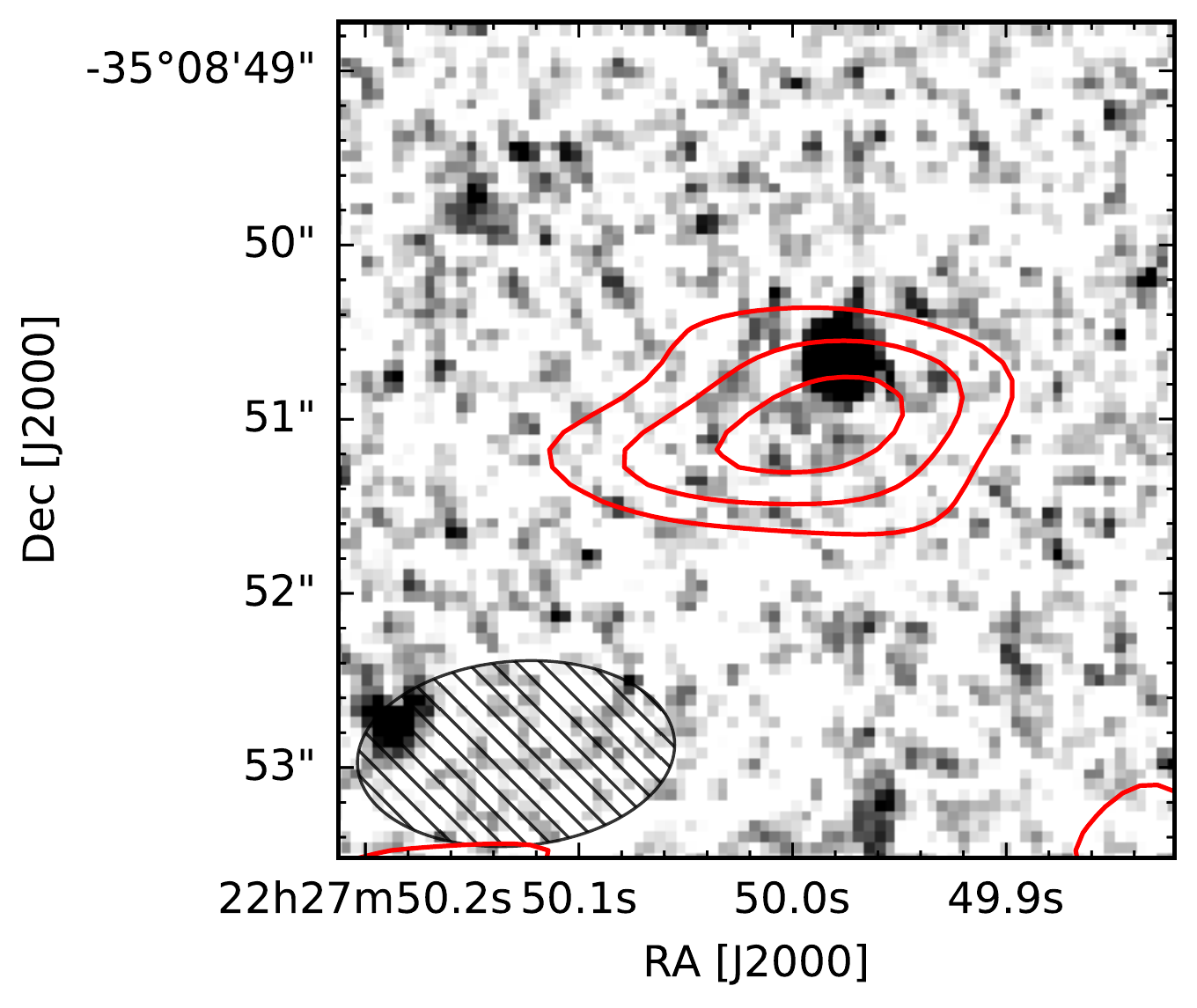}  %
\caption{Top: \cii\ flux map  of BDF2203 obtained integrating the ALMA cube over the velocity range between -60 km/s and 132 km/s. Red solid contours are at levels of 2$\sigma$,
3$\sigma$ and 4$\sigma$, where $\sigma=23$ mJy km/s. Dashed black contours indicate negative values at -3 and -2 times the noise level in the same map.
Bottom: zoom of the central 2.5\arcsec\ region around BDF2203. \cii\ flux map contours are over imposed on the NIR image. Red contours are at the same levels
as in the top panel. The ALMA synthesised beam  in the lower left corner in both panels. }
 \label{fig:mapBdf}
\end{figure}

The \cii\ emission is undetected in all but one of
these sources, that is, BDF2203.
Figure~\ref{fig:appFig1} shows the spectrum of the detected \cii\ emission,
with a spectral rebinning of 40 km s$^{-1}$, while the spectra of the non-detections are shown in the Appendix (Figure~\ref{fig:appFig2}).

For the four non-detections, we assume a full-width-at-half-maximum of
FWHM = 100 km s$^{-1}$, which is consistent with \cii\ line widths observed in other $z>6$ galaxies
\citep[e.g.][]{Pentericci:2016,Carniani:2017a, Carniani:2017}, and we infer  3$\sigma$ upper limits on the \cii\ luminosity (Table~\ref{tab:appTab1}). 

The redshift of the  \cii\ emission detected in BDF2203  is $z_{\rm [CII]}=6.1224\pm0.0005$, which is in agreement
with that inferred from Ly$\alpha$ ($z_{\rm Ly\alpha}= 6.12\pm0.03$). 
By fitting a 1D Gaussian profile to the \cii\ line we estimate a FWHM=$150\pm50$ km/s that is similar to those estimated in high-$z$ \cii-emitting galaxies \citep[FWHM=50-250 km/s;][]{Willott:2015, Pentericci:2016, Matthee:2017, Carniani:2017}.
The flux map of the \cii\ line, extracted with a spectral width of 200 km/s, is shown in  Figure~\ref{fig:mapBdf}.
The emission is detected in the map with a S/N=5 and has an integrated flux density of $140\pm35$ mJy km/s, which
corresponds to \lcii$=(12.5\pm2.5)\times10^{7}$~\lsun\ at  z=6.12. By fitting a 2D gaussian profile to the flux map, we
measure a  size  of ${\rm(2.3\pm0.5)\arcsec\times(1.2\pm0.2)\arcsec}$ with ${\rm PA=88\deg\pm8\deg}$ and a beam-deconvolved size of  ${\rm(1.4\pm0.8)\arcsec\times(0.5\pm0.4)\arcsec}$ with a ${\rm PA=80\deg\pm20\deg}$. The \cii\ emission is thus marginally resolved and has a diameter size of $\sim5$~kpc. The \cii\ flux map overlaps with the rest-frame UV emission that
has an extension of about 1 kpc. The centroids  are separated by 0.25\arcsec. However, given the ALMA beam size
($1.90\arcsec\ \times1.11\arcsec$), such positional offset is consistent with the positional uncertainties $\Delta\theta\approx\frac{<beam \ size>}{S/N}\sim0.2\arcsec-0.4\arcsec$. The low angular resolution and sensitivity of current observations are not sufficient to assess the morphology of the FIR line emission.


\subsection{SFR estimates and morphology analysis}
\label{sec:SFRestimates}

The continuum emission at rest-frame wavelengths around 158 $\mu$m is associated to thermal emission from dust heated by
the UV emission of young stellar populations in galaxies.
The continuum emission is detected only in four sources within the selected sample. When compared with the typical UV-IR SED of galaxies, the
weak rest-frame far-IR continuum indicates that these galaxies are on average characterised by low dust masses.

The spectral energy distribution (SED) of the thermal dust emission can be modelled with a greybody with dust
temperature \Td\ and spectral index emissivity $\beta$.  However, one single photometric measurement  is not sufficient
to perform the SED fitting and constrain the two free parameters. We thus assume \Td\ = 30K and  $\beta$ = 1.5, which are
consistent with those observed in local dwarf galaxies \citep{Ota:2014}, to estimate  the far-infrared (FIR) luminosity from the ALMA continuum observations. For those galaxies that are not detected in ALMA continuum images we infer a 3$\sigma$ upper limit on \lfir. 
We note that the FIR emission strongly depends on the assumed dust temperature,  yielding to a luminosity uncertainty of $\Delta \log({\rm L_{FIR}}) = 0.6 $ \citep{Faisst:2017a}.

For each galaxy we can infer the star-formation rate from both the UV (SFR$_{\rm UV}$) and  
 FIR (SFR$_{\rm FIR}$) emission by adopting the calibrations presented in \cite{Kennicutt:2012}.
Excluding HZ9 and HZ10, all  sources in the  sample have ${\rm SFR_{\rm FIR}/SFR_{\rm UV} \lessapprox 1}$ with an average value of ${\rm SFR_{\rm FIR}/SFR_{\rm UV}}\approx0.6$ . Given that most of the FIR luminosity estimates are $3\sigma$
upper limits, the average SFR$_{\rm FIR}$/SFR$_{\rm UV}$ is actually much lower than 0.6.

Given the small contribution of SFR$_{\rm FIR}$ (either in the IR-detected sources or those with upper limits), we assume total ${\rm SFR \approx SFR_{\rm UV}}$ with no dust correction for those galaxies without continuum detection. While we infer total ${\rm SFR = SFR_{\rm UV}+SFR_{\rm FIR}}$ for those galaxies revealing continuum emission at 158 $\mu$m. 

In order to assess the multi-clump morphology of the \cii\ emission in our sources, we perform a kinematical analysis on the retrieved ALMA cube, extracting channel
maps at different velocities relative to the redshift of the galaxies. This analysis enables us to disentangle the emission of complex systems having multiple
components at different velocities \citep[e.g.][]{Carniani:2013, Riechers:2014}. We also estimate the size of the
various UV and \cii\ emission by fitting a 2D elliptical Gaussian profile to emission maps and by taking into account
the angular resolution of the observations.

The properties of the \cii, FIR and UV emission, such as redshift, flux density, luminosity, SFR, and radius, are reported in Table~\ref{tab:table1}


\section{Multi-components systems}\label{sec:Multi-component systems}

\begin{figure*}
\centering
\includegraphics[width=2\columnwidth]{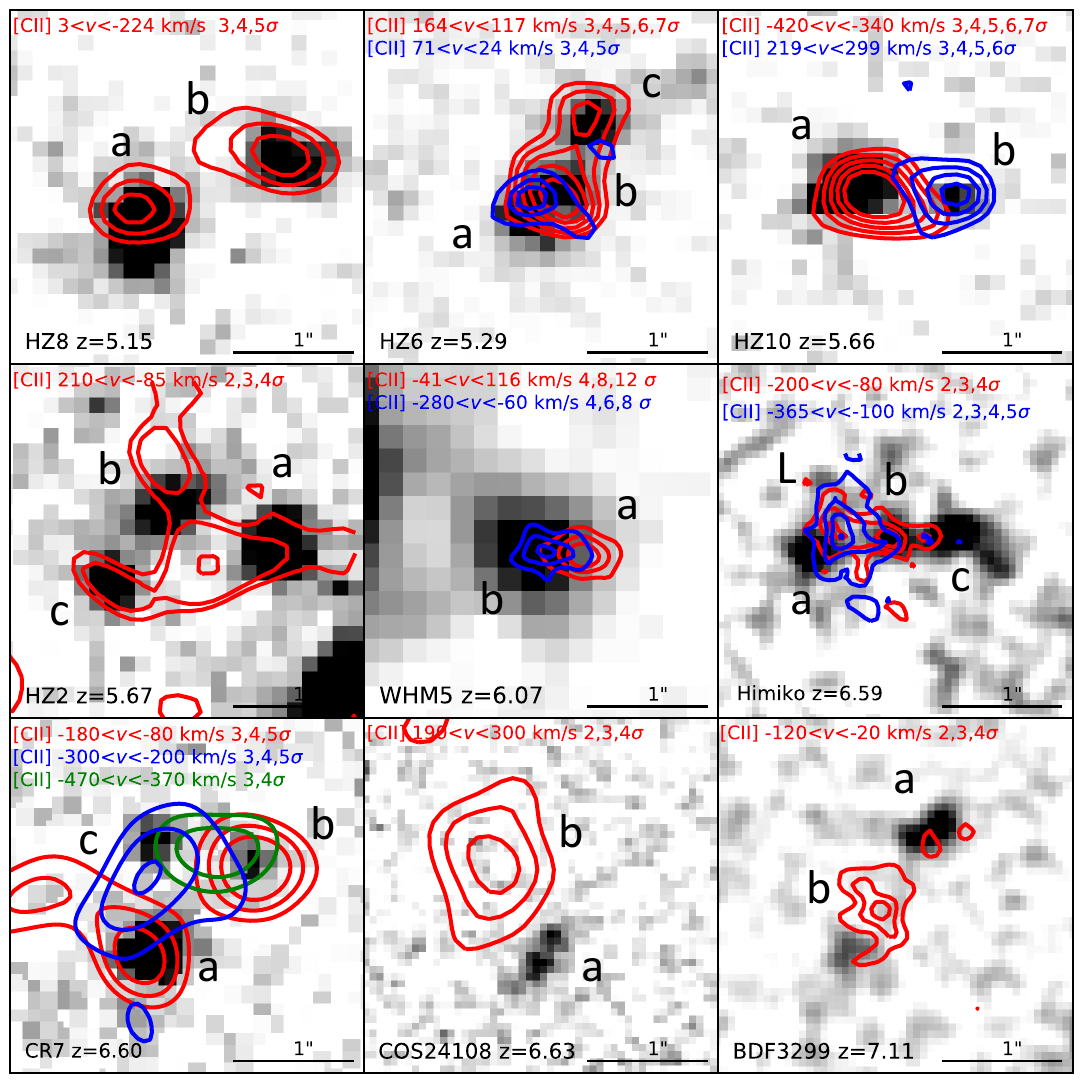}
 \caption{Rest-frame UV images of the nine  $z>5$ star-forming galaxies showing multi-clump morphology in \cii\ and/or rest-frame UV emission. The galaxies are ordered by redshift
 and each stamp is 3\arcsec\ on the side, with North to the top and East to the left. The red, blue, and green contours show the  \cii\ channel maps at different velocity intervals. 
 The levels of the contours and channel maps velocities are indicated in the legends. The properties of each galaxy are reported in Table~\ref{tab:table1}.}
 \label{fig:fig1}
\end{figure*}

\cite{Carniani:2017a}  discuss the origin  of the positional and spectral offsets between optical emission and FIR lines   observed in  high-$z$ systems, suggesting that the observational properties can arise from  distinct regions (or components) of galaxies. In order to understand the nature of these offsets, we have re-analysed  ALMA \cii\ observations for those galaxies showing \cii\ detections and have performed a morphology analysis as discussed in Section~\ref{sec:SFRestimates}.
We have found a clear multi-clump morphology in nine out of 21 $z>5$ galaxies having \cii\ detections (Table~\ref{tab:table1}). 
The UV rest-frame images and \cii\ maps are shown in Figure~\ref{fig:fig1}, sorted by redshift.
The stamps are 3\arcsec\ across, or $\sim$17.5~kpc at the average redshift of this sample ($<z>= 6$), and show
the contours of the \cii\ maps, obtained from our analysis, superimposed on 
the rest-frame UV emission in grayscale. 
%

New {\it HST} near-IR images of HZ6, HZ8, and HZ10 have already been presented in \cite{Barisic:2017} and \cite{Faisst:2017a} revealing multi-component structures.
The location of the individual rest-frame UV clumps  is consistent with the  peak positions  of the \cii\ emission extracted at different velocities relative to the
redshift of the brightest component (which is labelled ``a'' in all stamps).  
The \cii\ emission detected in all individual clumps has a level of significance\footnote{$\sigma$ is the rms of the channel map in which we detect the \cii\ emission} higher than 5$\sigma$ and it is spatially  resolved (see Table~\ref{tab:table1}). 
The channel map analysis confirms that HZ8b and HZ10b (Hz8W and Hz10W in previous works) are at the same redshift of
HZ8a and HZ10a, respectively.  We note that the kinematic properties of HZ10 are consistent with the analysis reported by \cite{Jones:2017a}, who claim that the velocity gradient observed in this galaxy  matches a merger scenario rather than a rotating gas disk model.

The deeper {\it HST} observations reveal also two faint companions close to the HZ2 galaxy within a projected distance of $\sim$6~kpc.
These sources are not discussed in previous studies \citep{Capak:2015,Barisic:2017}, since their redshifts were not
fully spectroscopically confirmed by the \cii\ line. 
However, a detailed kinematic investigation of the carbon line shows an extended emission with a morphology consistent with the rest-frame UV emission. 
Although the two faint companions are detected with a low level of significance ($\sim3.5\sigma$), the match between the three UV peaks and the \cii\ emission supports the reliability of the ALMA detections. 
The three sources, dubbed in this work as HZ2a, HZ2b and HZ2c, are  at the same redshift and form a multi-component system similar to that observed in HZ6.
The angular resolution and the low sensitivity of current ALMA observations are not sufficient to spatially resolve the \cii\ emission in HZ2b and HZ2c.
HZ2a is instead spatially resolved, with a diameter of 2.5 kpc.

An additional  star-forming galaxy with a complex morphology is WHM5 at $z=6.0695$, which appears to consist of multiple components in \cii: a compact  source,  seen also in dust emission, and an extended component at the location of the rest-frame UV emission \citep{Willott:2015a, Jones:2017}. 
The two components are separated by a projected distance of $\sim3$ kpc and a velocity of $\sim200$ km/s. Both \cii\
emission components are spatially resolved ($\sim$1.3 and $\sim$2.8 kpc) in the ALMA observations with an angular resolution of 0.3\arcsec\ \citep{Jones:2017}.

At higher redshift, the presence of multiple \cii\ components has recently been reported in CR7 at $z$=6.6  by
\cite{Sobral:2015,Sobral:2017} and \cite{Matthee:2015,Matthee:2017}. Three out of four detected \cii\ clumps  coincide
with the location of UV clumps. In two cases the \cii\ emission  is spatially resolved with a radius of $\sim3-3.8$ kpc (see \citealt{Matthee:2017} for details).

An additional multi-clump galaxy observed with ALMA is Himiko at $z\sim6.695$ \citep{Ouchi:2013}. The rest-frame UV image (see Figure~3 of \citealt{Ouchi:2013})
reveals that the galaxy comprises three sub-components with SFR spanning in range between 5 and 8 \sfr. The projected distance between the sources is of about 3-7 kpc.
kpc. Out of the three sub-components have a \lya\ EW=68 \AA\, while the other two have EW less than 8 \AA. Although early observations had reported non-detections
of \cii\ in this source, recently \cite{Carniani:2017} have reported a clear detection with extended/multi-clump morphology. The primary \cii\ emission is
coincident with the Ly$\alpha$ peak, while the UV clumps are much weaker in \cii\ or even undetected. More generally the extended \cii\ emission does not resembles the UV clumpy distribution.

For the remaining systems (COS24108 and BDF3299) showing a clear positional offsets between \cii\ and UV emission in Figure~\ref{fig:fig1}, we cannot speculate much more than what has been already done in  previous works due to the lack of deeper ALMA and/or {\it HST} observations.  
As discussed by \citealt{Carniani:2017a},  these offsets are certainly associated with physically distinct sub-components, and the \cii\ clumps with no
UV counterpart may either be tracing star forming
regions that are heavily obscured at UV wavelengths \citep{Katz:2017}, or associated with accreting/ejected gas.


\subsection{On the nature of the kpc-scale sub-components} 

The deeper \cii\ and  rest-frame UV observations have unveiled the real multi-component nature of nine star-forming galaxies at $z>5$ further highlighting and
(partly) explaining
positional offsets between UV and \cii\ emission in previous studies \citep{Capak:2015,Maiolino:2015,Willott:2015,Pentericci:2016,Carniani:2017a}. These results
suggest that future evidence of displaced FIR line and UV emission should be not ignored since it generally reveals the
presence of sub-components with different physical properties.

Such sub-components can be ascribed to either satellites in the process of accreting \citep{Pallottini:2017a} or clumps ejected by past galactic outflows \citep{Gallerani:2016}.
However, in many cases both the SFR$_{\rm UV}$ and the size of the various sub-components are comparable to those estimated for
the central galaxies, hence suggesting a
major merger scenario in many cases.
Future ALMA observations with higher resolution and sensitivity are necessary to detect further sub-components in these and other systems and to perform a detailed dynamical analysis, which allow us to to estimate the dynamic mass and assess the nature of sub-components.



\section{The \lcii\ - SFR relation at z=5--7}\label{sec:dispersion}

A tight relation between the \cii\ luminosity and the global SFR is seen in local galaxy observations, at least
when excluding extreme (ULIRG-like) cases \citep{De-Looze:2014, Kapala:2015, Herrera-Camus:2015}.
This finding makes the \cii\ line a promising tool to investigate the properties of early galaxies and to
trace their star formation. 
However,  the behaviour of the \cii\ line emission at $z>5$ seems to be more complex than observed in the local Universe.  
Previous studies have shown that only a fraction of \cii\ detections of early galaxies agree with the local relation, while most
high-$z$ galaxies are broadly scattered, with claims that most of them are \cii-deficient relative to the local relation. 
However, most of previous high-$z$ studies classified multi-component systems as single objects in the \lcii-SF diagram.
If we associate each clump and/or galaxy with its proper UV counterparts (or lack thereof), then the resulting location on the \lcii-SFR diagram changes significantly for these objects

\begin{figure*}
\centering
\includegraphics[width=1.6\columnwidth]{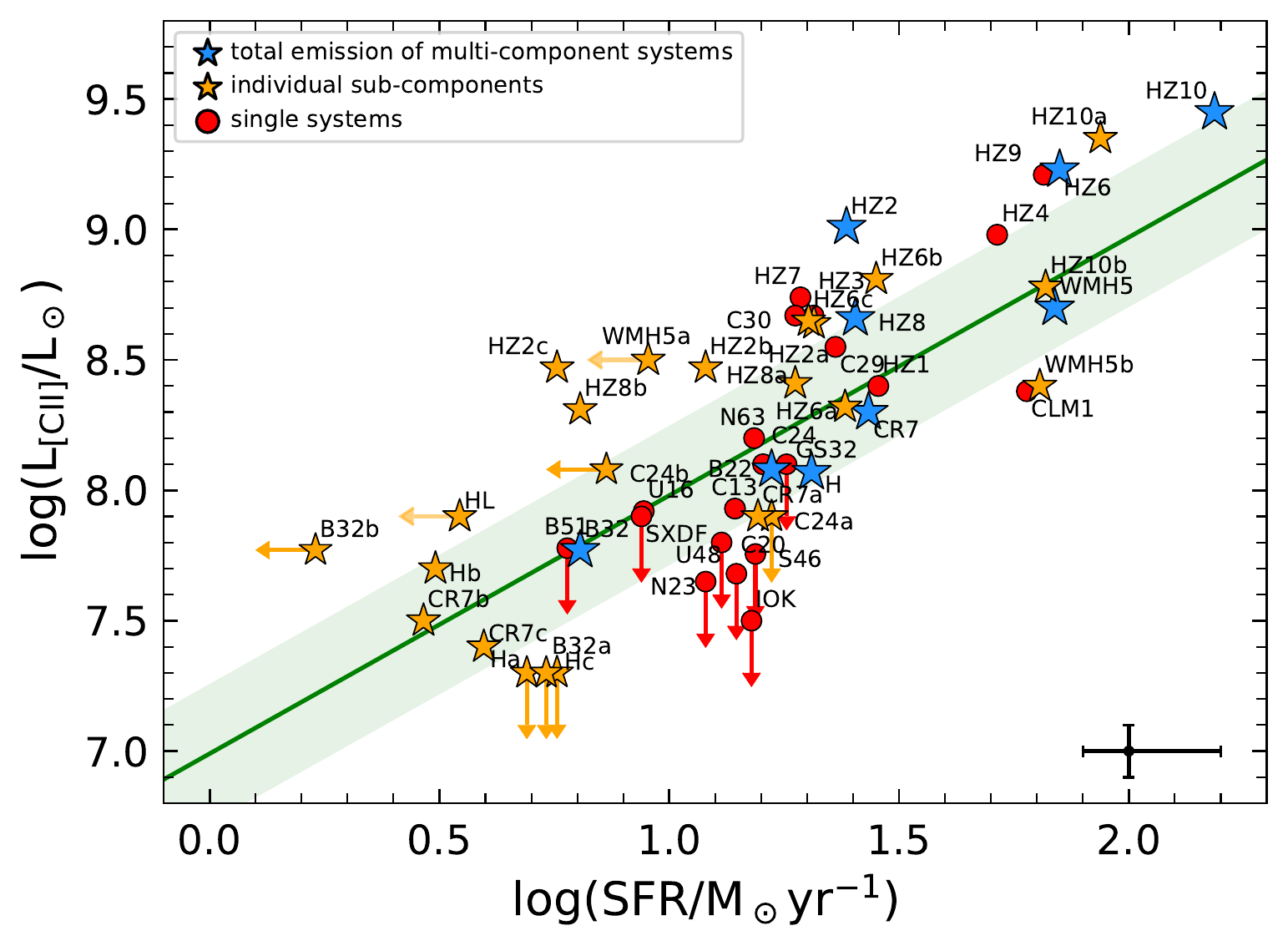}\\\ \\
 \caption{\lcii\ as a function of SFR at $z=5-7$. Blue stars indicate the global SFR and \cii\ luminosity for system having multiple sub-components,
 as discussed in Section~\ref{sec:Multi-component systems}; in these cases the location of the individual sub-components are shown with yellow stars. 
Red circles show the location of the remaining high-$z$ galaxies (i.e. single systems) listed in Table~\ref{tab:table1}, that do not exhibit  any positional offsets and/or disturbed morphology.
In the bottom-right of the figure we show an error bar that is representative for the whole sample.
The green line is the local relation for local star-forming galaxies \citep{De-Looze:2014}.}
 \label{fig:fig2}
\end{figure*}

Figure~\ref{fig:fig2} shows  \lcii\ as a function of SFR. The green line illustrates the local relation obtained
by \cite{De-Looze:2014} and its dispersion is given by the
shaded area. Results for $z>5$ galaxies, as listed in Table~\ref{tab:table1}, are shown with various symbols in
Figure~\ref{fig:fig2}. 
The SFR estimation for the $z>5$ galaxies (and their sub-components) is discussed in Section~\ref{sec:SFRestimates}.
The multiple-component objects (HZ2, HZ6, HZ8, HZ10, WHM5, Himiko, CR7, COS24108, and BDF3299) are split into several individual components with their own SFRs and \lcii. 
The nine complex systems discussed in Section~\ref{sec:Multi-component systems} are broken into 20 sub-components distributed on different
regions of the \lcii-SFR plane. The location of the individual subcomponents is indicated with yellow stars, while the location of these
systems by integrating the whole \cii\ and UV emission (i.e. ignoring that these are actually composed of different subsystems) is
indicated with blue stars.
The four new \cii\ non-detections presented  in Section~\ref{sec:additional} fall below the local relation, while the \lcii\ for BDF2203 places this galaxy along the \cite{De-Looze:2014} relation.

%
Once the association between \cii\ emission and optical-UV counterparts is properly done,
we find that the resulting distribution occupies a large area of the  \lcii-SFR plot with a large scatter  both above and below the local relation.
About 19 objects of the total sample  are in agreement within 1 sigma with the local relation, but the remaining 24 systems have deviations, either above or below the relation, up to 3 sigma.

\begin{figure}
\centering
\includegraphics[width=1\columnwidth]{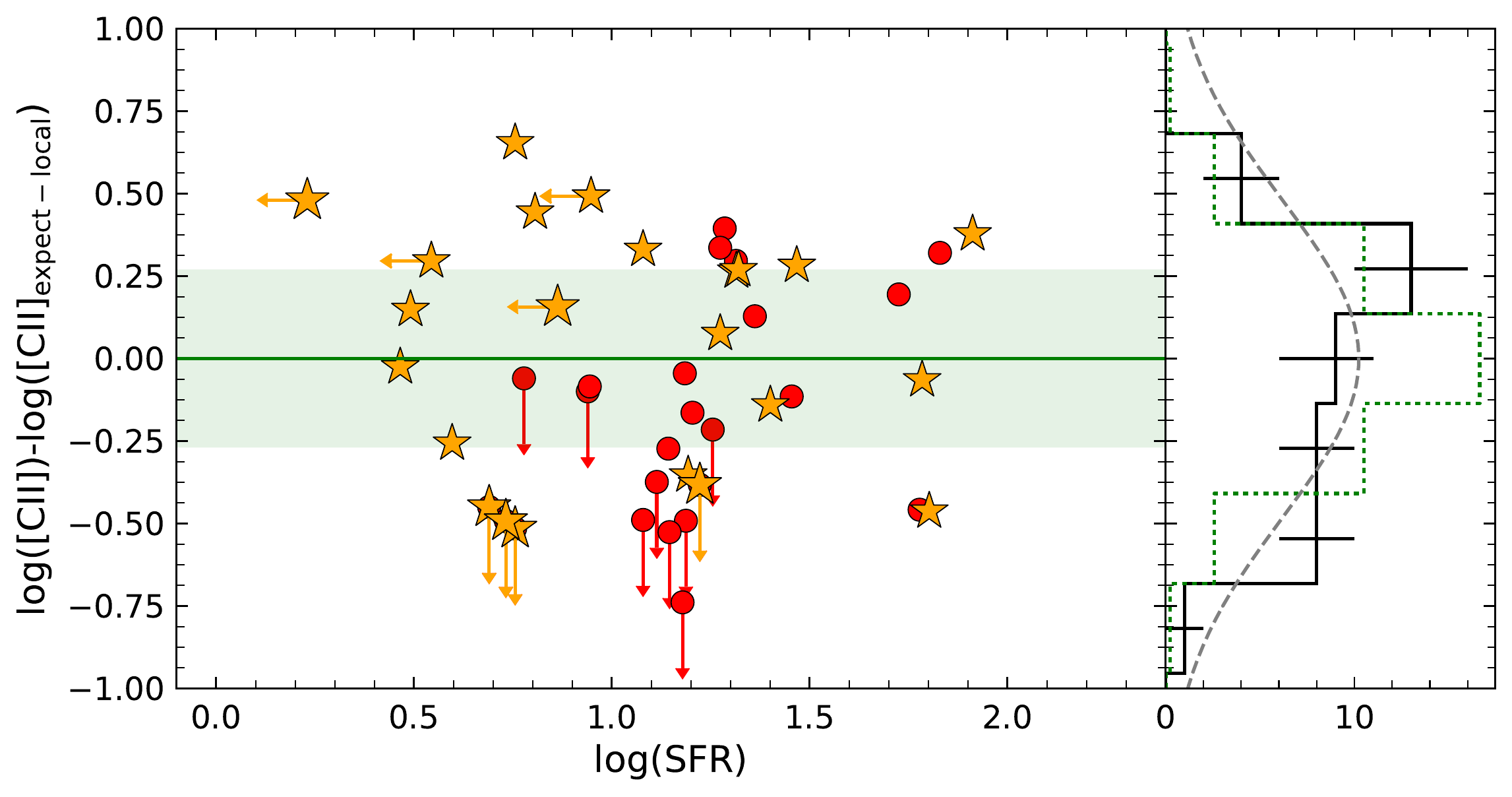}
\caption{Deviations from the local \lcii-SFR relation as a function of SFR. The 1$\sigma$  dispersion of the local relation is indicated with the shaded green region.
The distribution of the offsets is shown in the right panel. The black curve represents the Log(\cii)-Log(\cii$_{\rm expect-local}$) distribution from our sample, while the dotted
green histogram is the distribution of the local relation. The dashed grey line shows a Gaussian fit to the high-$z$ distribution.}
 \label{fig:fig2b}
\end{figure}

In order to quantify the \lcii-SFR offset of the high-$z$ sample from the relation found in the local population, we investigate the distribution of offsets relative to the local
relation. More specifically,
for each galaxy we calculate the offset from the relation as $\Delta$Log(\cii)=Log(\cii)-Log(\cii$_{\rm expect-local}$), where Log(\cii$_{\rm expect-local}$) is the \cii\ luminosity
expected from the local relation according to the SFR measured in the galaxy or sub-component.
The result of this distribution is shown in the left panel of Figure~\ref{fig:fig2b}, while the right panel shows the distribution of the offsets.  
In contrast with some previous claims based on fewer targets, the $\Delta$Log(\cii) distribution, which includes both detections and upper limits, does not exhibit  any clear shift relative
to the local relation (and whose distribution is shown with the dotted green histogram.
The number of objects below (23)  and above (20)  the \lcii-SFR relation are comparable. 
 However, the dispersion of the
is 0.48$\pm$0.07, which is about two times larger than the uncertainty reported by \cite{De-Looze:2014} for the local relation. Such larger dispersion may be associated to the presence of kpc-scale sub-components that are not common in the local Universe. However, we will discuss the possible origin of this dispersion in the next sections.

\section{The relation between \lya\ EW and \cii\ emission}\label{sec:lyaEW}

It is well known that \lya\ emission depends on the level of ionizing photons produced by star formation (or AGN activity) and radiative transfer effects in the ISM.
Models and observations suggest that the \lya\ EW increases with decreasing  metallicity and dust content \citep{Raiter:2010, Song:2014}. 
Since the \cii\ emission is sensitive  on the ISM properties as well, and in particular the ISM heating through photoelectric ejection from dust grains,
we expect a relation between the \cii\ luminosity and the \lya\ strength \citep{Harikane:2017}.

\begin{figure}
\centering
\includegraphics[width=1\columnwidth]{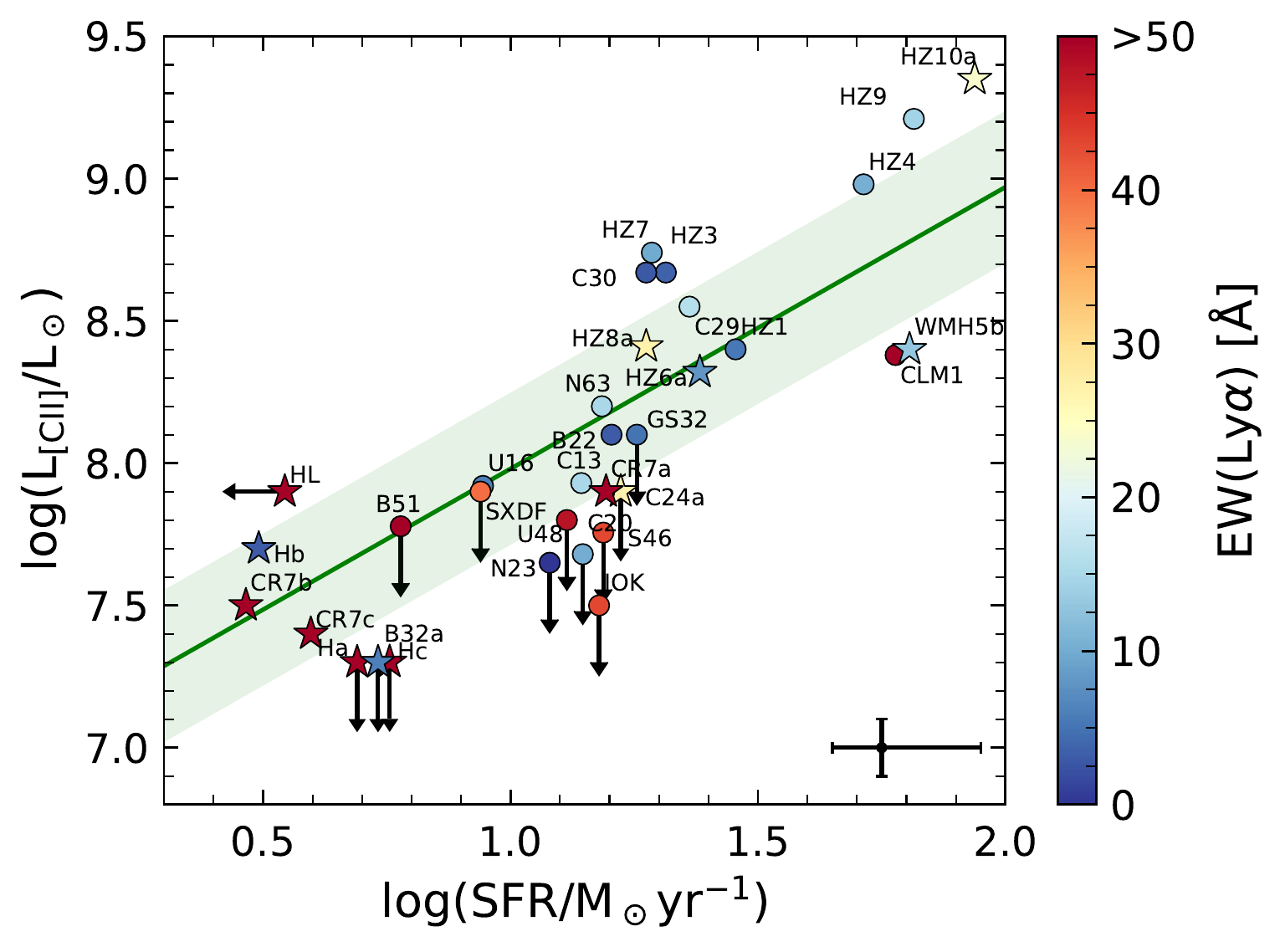}
\caption{  \lcii\ as a function of SFR at $z=5-7$. 
Stars and circles represent sub-components and individual galaxies, respectively.
Symbols are colour-coded according to  their  \lya\ EWs (not corrected for the inter-galactic medium  absorption), as indicated on the colour bar on the right.
In the bottom-right corner we show an error bar that is representative for the whole sample.
The green line are the local relation for local star-forming galaxies \citep{De-Looze:2014} while its dispersion  is indicated by the shaded green region.}
 \label{fig:EW}
\end{figure}

\begin{figure}
\centering
\includegraphics[width=1\columnwidth]{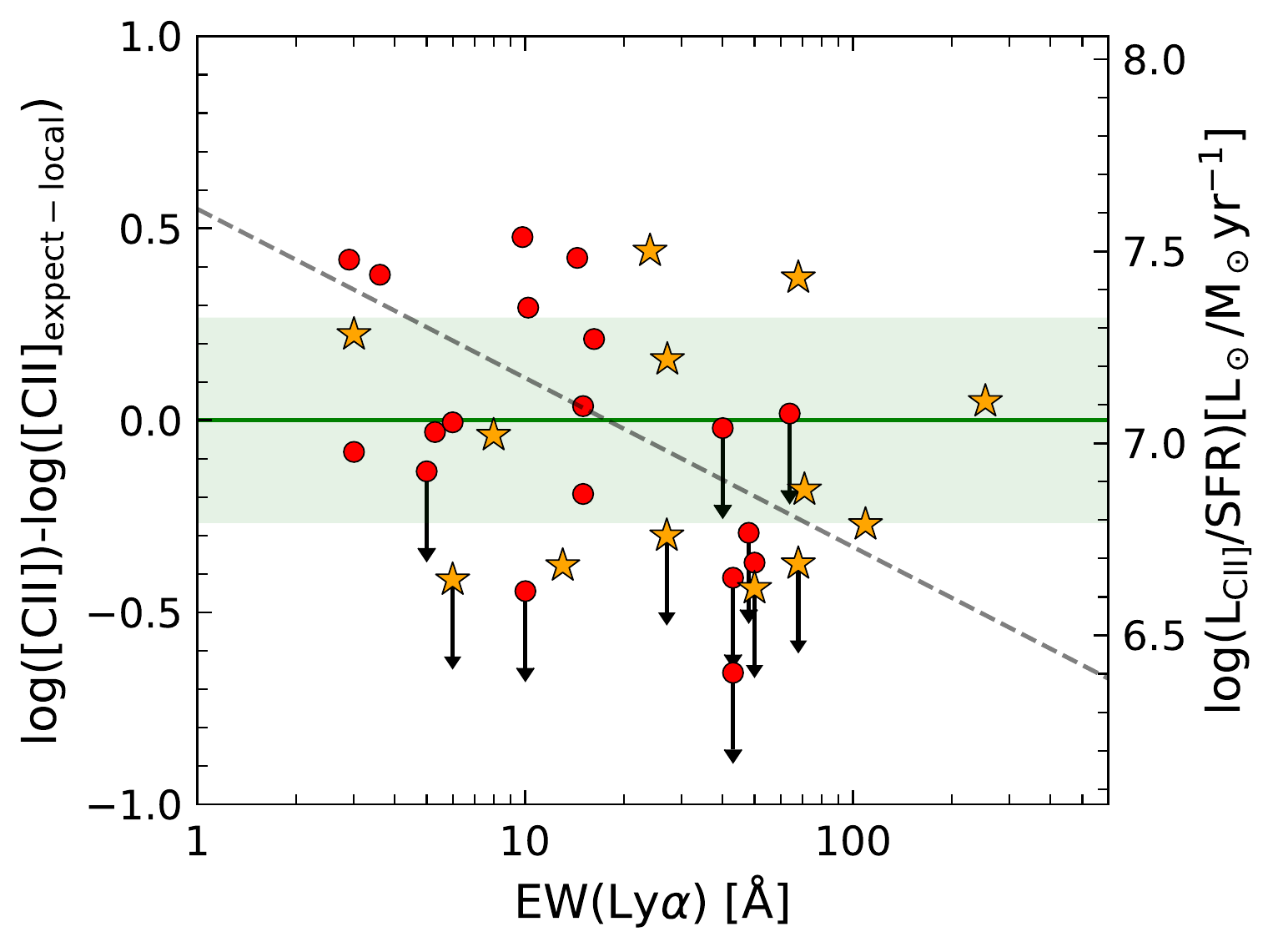}
\includegraphics[width=1\columnwidth]{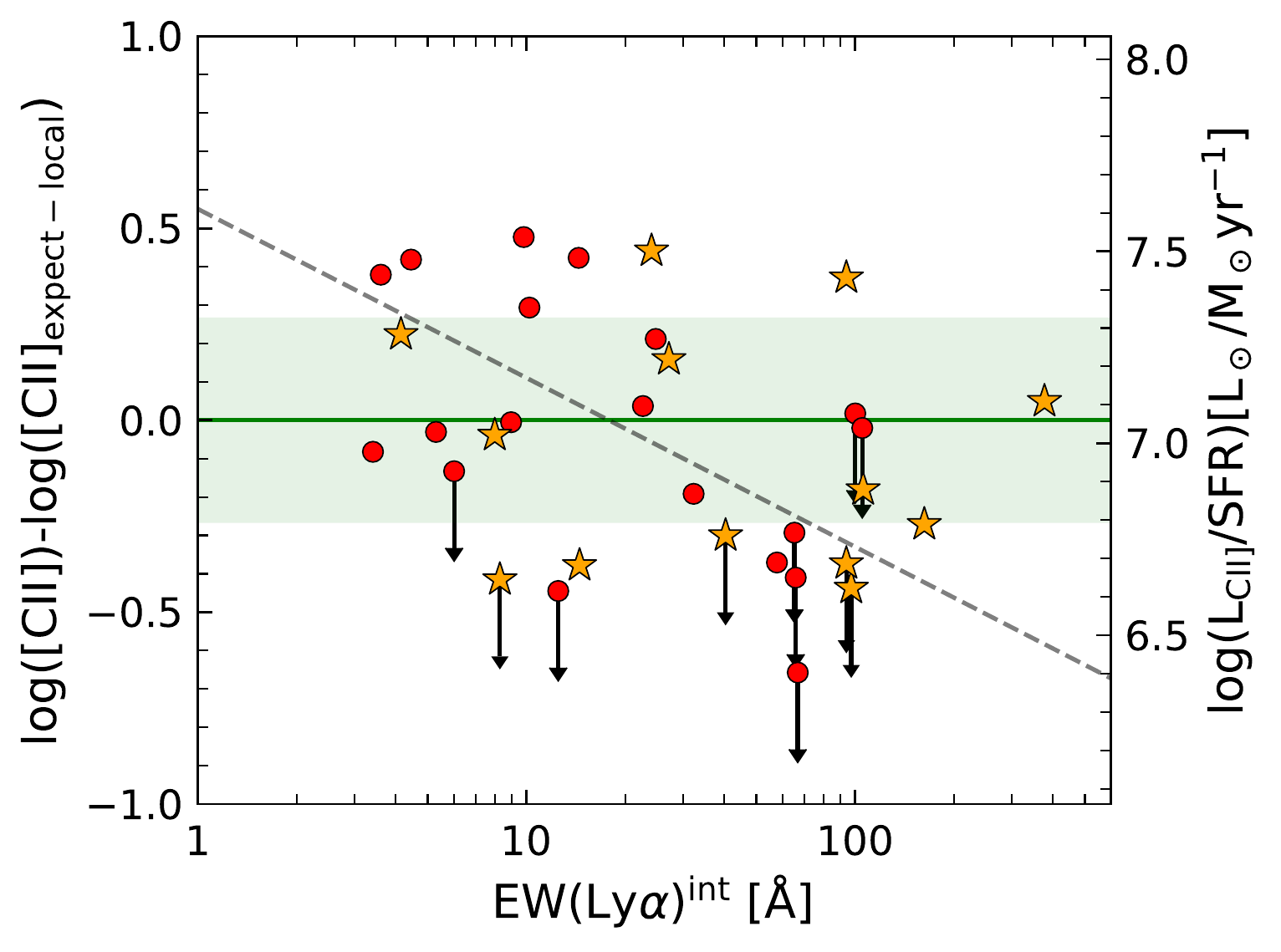}
\caption{ Top panel: Offset from the local \lcii-SFR relation  as a function of EW(Ly$\alpha$).  The right hand axis shows the \cii/SFR ratio corresponding to the
deviations from the local \lcii-SFR relation.
The 1$\sigma$  dispersion of the local relation is indicated by the shaded green region. Star and circles indicate sub-components and individual galaxies, respectively.
  The grey dashed relationship shows our linear fit (see text).
  Bottom panel: Same as the top panel but the EW(Ly$\alpha$)$^{\rm int}$ has been corrected for the IGM absorption by following \citealt{Harikane:2017}.}
 \label{fig:EW2}
\end{figure}

In Figure~\ref{fig:EW}  we show  \lcii\ as function of SFR by colour-coding the different symbols according to their \lya\ EWs (not corrected for the inter-galactic medium  absorption). We include 
only those sub-components and galaxies having a \lya\ EW measurement.
There is a weak tendency for galaxies with high EW(Ly$\alpha$) to lie below the local \lcii-SFR relation, and vice-versa.
This is shown better in Figure~\ref{fig:EW2} where the offset from the local \lcii-SFR relation
is plotted as a function of EW(\lya). Note that, since the slope of the local relation is one, plotting the deviation from the local relation
($\Delta\log$(\cii)=log(\cii)-log(\cii$_{\rm expect-local}$)) is equivalent to plotting the ratio between \cii\ luminosity and SFR, i.e. L$_{\rm [CII]}$/SFR, which is indeed given on the
right hand axis of  Figure~\ref{fig:EW}.
Although there is a large dispersion, there is a tentative indication that the offset from the local relation (hence L$_{\rm [CII]}$/SFR) anti-correlates with \lya\ EW. A linear
fit gives
$$
{\rm \Delta \log([C\textsc{ii}]) = (0.55\pm0.20)-(0.44\pm0.15)\log(EW(Ly\alpha)) } 
$$
and the dispersion around this best-fit is 0.25~dex.

The result does not change significantly if we attempt to correct the EW(\lya) for IGM absorption, by following the prescription given by \cite{Harikane:2017}.
In this case the relation is shown in the bottom panel of Figure~\ref{fig:EW} and the resulting best-fit linear relation is
$$
{\rm \Delta \log([C\textsc{ii}]) = (0.56\pm0.20)-(0.41\pm0.14)\log(EW(Ly\alpha)^{int})  \ \  (0.30 {\it dex})}
$$
and the dispersion around this best-fit is even larger, 0.3~dex.

\cite{Harikane:2017} find a relation between L$_{\rm [CII]}$/SFR and EW(\lya) steeper than ours, though consistent within errors. The steeper relation found by
\cite{Harikane:2017} is probably associated with the fact that in their work they combine the global properties of galaxies and do not extract the subcomponents.

Some anti-correlation between deviation from the local \lcii-SFR relation and EW(\lya) (or, equivalently, between \lcii-SFR and EW(\lya)$^{\rm int}$) is expected
from the dependence of these quantities from the metallicity, either directly
or through the associated dust content, as already predicted by some models \citep[e.g.][]{Vallini:2015,Pallottini:2017, Matthee:2017}.
Indeed, lower metallicity implies lower amount of carbon available for cooling, but also less dust content. Indeed, since the heating of PDRs occurs primarily
through photoelectric effect on dust grains, the lower is the dust content the lower is the heating efficiency of the gas in the PDR, hence the lower is the
emissions of the \cii\ cooling line. On the other hand, the lower the dust content the lower is the absorption of the Ly$\alpha$ resonant line, hence the higher
is the EW(\lya).

\section{Spatially resolved \lcii--SFR relation}\label{sec:spatially}

In the previous sections, we show that the \lcii-SFR relation at $z>5$ has a intrinsic dispersion larger than observed in the local Universe.
Such a large scatter suggests that the \cii\ luminosity may not be good tracer of the SFR at least at early epochs. 

Recent spatially resolved studies have claimed that the \cii-SFR relation is better behaved in terms  of SFR surface density and \cii\ surface brightness  than in global  proprieties (\lcii\ and SFR), since the surface brightness calibration is more closely related to the local UV field  \citep{Herrera-Camus:2015, Smith:2017}.
It is thus worth to analyse the relation \ecii-\esfr\ at $z>5$. 

In this section we compare the spatial extension of the \cii\ and UV emission and, then, we investigate the correlation between \cii\ surface brightness and SFR surface density.

\subsection{Spatial extension of the \cii\ emission}\label{sec:spatialextension}

\begin{figure}
\centering
\includegraphics[width=1\columnwidth]{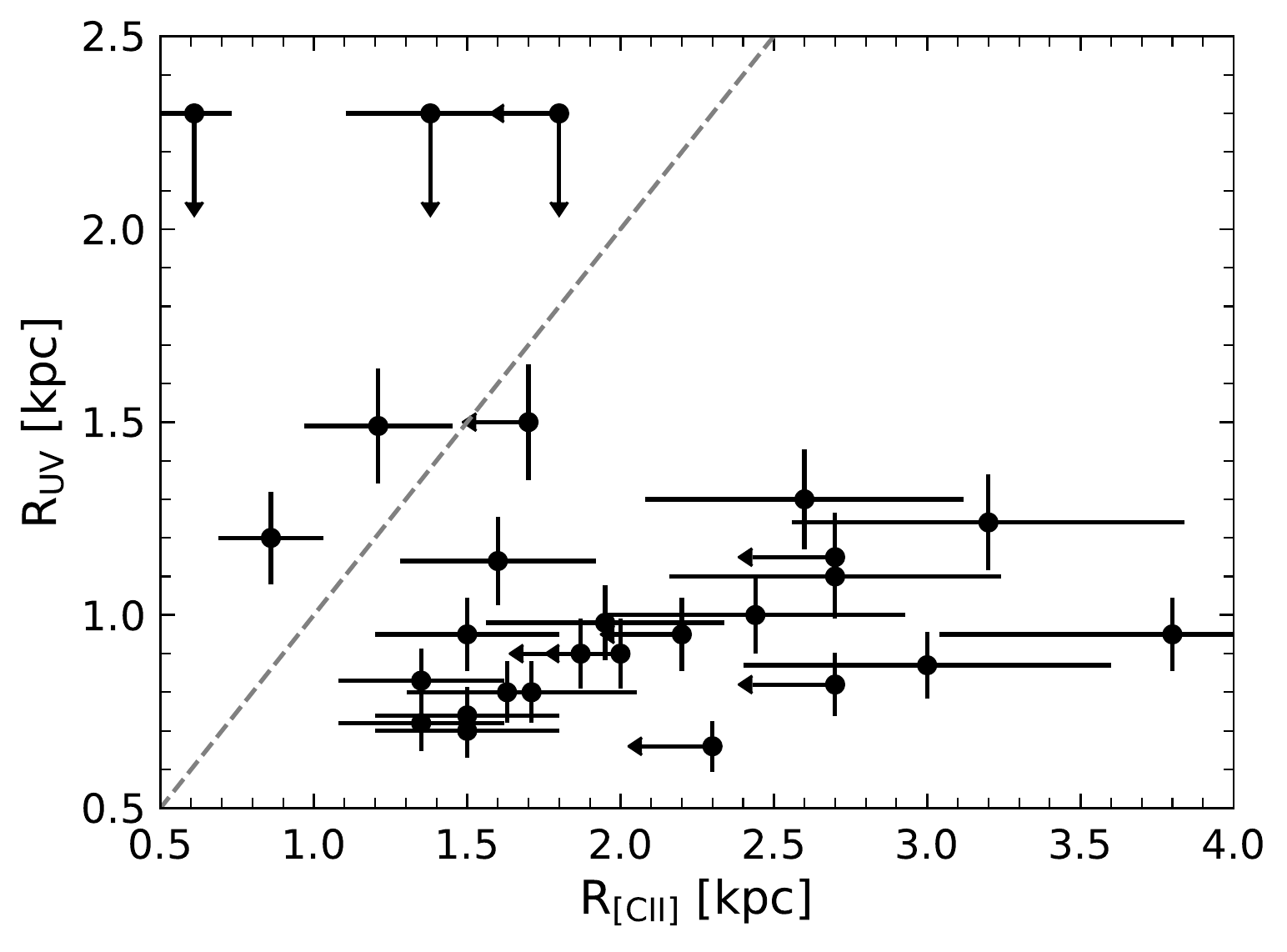}\\\ \\
\caption{Half-light radii of star formation regions, as measured from the (rest-frame) UV light, compared with the half-light radii of the associated \cii\ emission. The dashed line indicates the 1:1 relation.}
 \label{fig:fig_ext}
\end{figure}

Figure~\ref{fig:fig_ext} shows the extension of the \cii\ emission compared with the extension of star formation traced by the UV counterpart.
\cii\ emission is generally much more extended than the UV emission tracing unobscured star formation. This discrepancy may be partially
associated with observational effects. Indeed, while the high angular resolution of {\it HST} enables to resolve small clumps, it may have low
sensitivity to diffuse, extended emission. However, even by smoothing the UV images in the deepest observations available to us, we still
do not recover the extension observed in \cii. On the other hand, the resolution of the \cii\ observations may, in some case, smear out
clumps and result in an overall extended distribution. However, in many cases the ALMA observations achieve a resolution comparable, or even
higher, than {\it HST} at UV rest-frame wavelengths and, despite this, we measure clearly larger \cii\ sizes. Moreover, when high angular resolution
observation are used, these do reveal that a significant fraction of the \cii\ flux is resolved out on large scales \citep[see e.g. discussion
in ][]{Carniani:2017a}. In conclusion, we believe that the different sizes between \cii\ and UV emission are tracing truly different distribution
of the \cii\ and UV emission on different scales.

There could be various explanations for these differences. If the star formation associated with the \cii\ emission
on large scale is heavily obscured, the UV light does not trace this component \citep{Katz:2017}. This is certainly a possibility, although 
one might expect that the bulk of the obscuration should affect the central region more heavily than the outer parts, hence one would
expect the opposite trend. In alternative, the extended component of \cii\ may not be directly associated with star forming regions, but with
circumgalactic gas, either
in accretion or ejected by the galaxy, and which is illuminated by the strong radiation field produced by the galaxy.

\begin{figure}
\centering
\includegraphics[width=1\columnwidth]{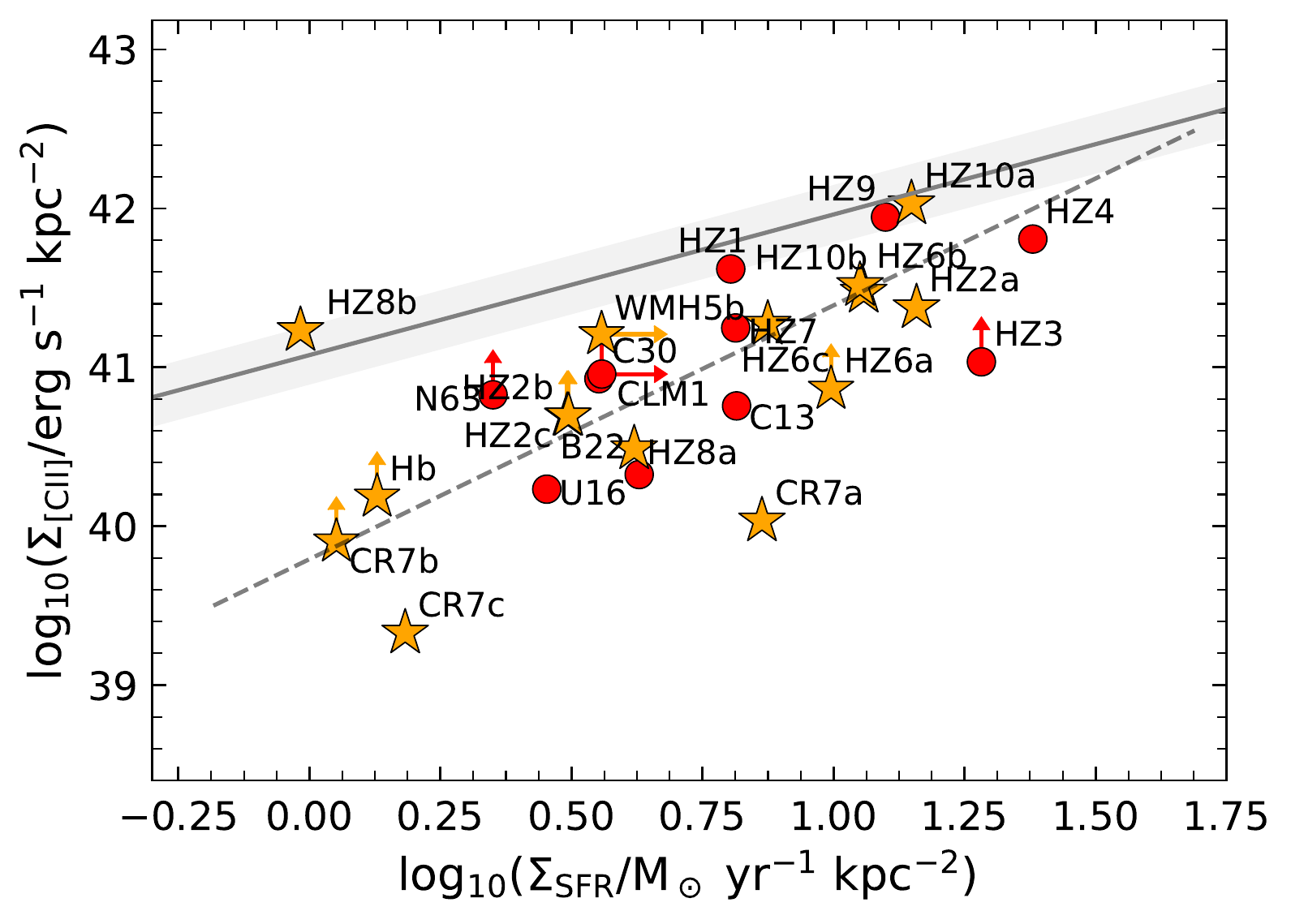}\\\ \\
\caption{\ecii vs \esfr \ for $z>5$ star-forming galaxies detected both in \cii\ and rest-frame UV. Symbols are as in Figure~\ref{fig:fig2}. The grey line indicates the local
relation by \citet{Herrera-Camus:2015}. The dashed line indicates the best linear fit on the $z>5$ sample.}
 \label{fig:fig3}
\end{figure}

\subsection{Surface brightness}\label{sec:surface brightness}\label{sec:surface}

Once we have measured the extension of the \cii\ emission and of the SFR regions, we can estimate the \cii\ surface brightness and SFR surface density of each sub-component and of
the individual sources detected in \cii\ and UV.
Figure~\ref{fig:fig3} shows the \ecii-\esfr\ relation, where we have included only those galaxies detected in both \cii\ and UV emission. 
Systems that are not spatially resolved in \cii\ (or UV) emission are indicated with lower limits. We also show the local relation by \cite{Herrera-Camus:2015} and its dispersion. 
In contrast to the \lcii-SFR diagram, there are no galaxies located significantly above the local relation, only a few galaxies
are located on the local relation, and most galaxies spread largely below the local relation. This is primarily due to the
large extension of the \cii\ emission in these high redshift systems, as discussed in the previous section.
By fitting the \ecii-\esfr\ measurements for our sample we obtain the following relation:
$$
\log(\Sigma_{\rm SFR}) = (0.63\pm0.11)\times(\log(\Sigma_{\rm [CII]})-(25\pm6)
$$
Part of the \ecii\ deficit may be ascribed to the metallicity of the gas.
Indeed a similar deviations have been  observed in local low-metallicity galaxies, in which the \ecii-\esfr\ calibration over-predicts the \ecii\
by up to a factor of six.
We note that \cite{Faisst:2017a} estimated a  metallicity of 12+log(O/H)$>$8.5 for HZ10 and HZ9, and  a metallicity of  12+log(O/H)$<$8.5 for HZ1, HZ2, and HZ4. 
The former agree with the local \ecii-\esfr\  relation while HZ1, HZ2, and HZ4 have low \ecii-\esfr\ ratio.
This interpretation is also supported by the fact that all CR7 clumps, which have  metallicities 12+log(O/H)$<8.2$ (see discussion in \citealt{Sobral:2017}) , are located well below the local relation.  
The \ecii\ deficit is also akin to the simulations by \cite{Pallottini:2017a} who investigated  the physical properties of a simulated galaxy at $z=6$, with metallicity
12+log(O/H)=8.35.  Such galaxy has a ${\rm \log(\Sigma_{\rm[CII]}/erg \ s^{-1} \ kpc^{-2})=40.797}$ and ${\rm \log(\Sigma_{\rm SFR}/M_\odot yr^{-1})= 1.027}$, which places the mock
galaxy below the  \ecii-\esfr\ relation found for local star-formation galaxies.

\begin{figure}
\centering
\includegraphics[width=1\columnwidth]{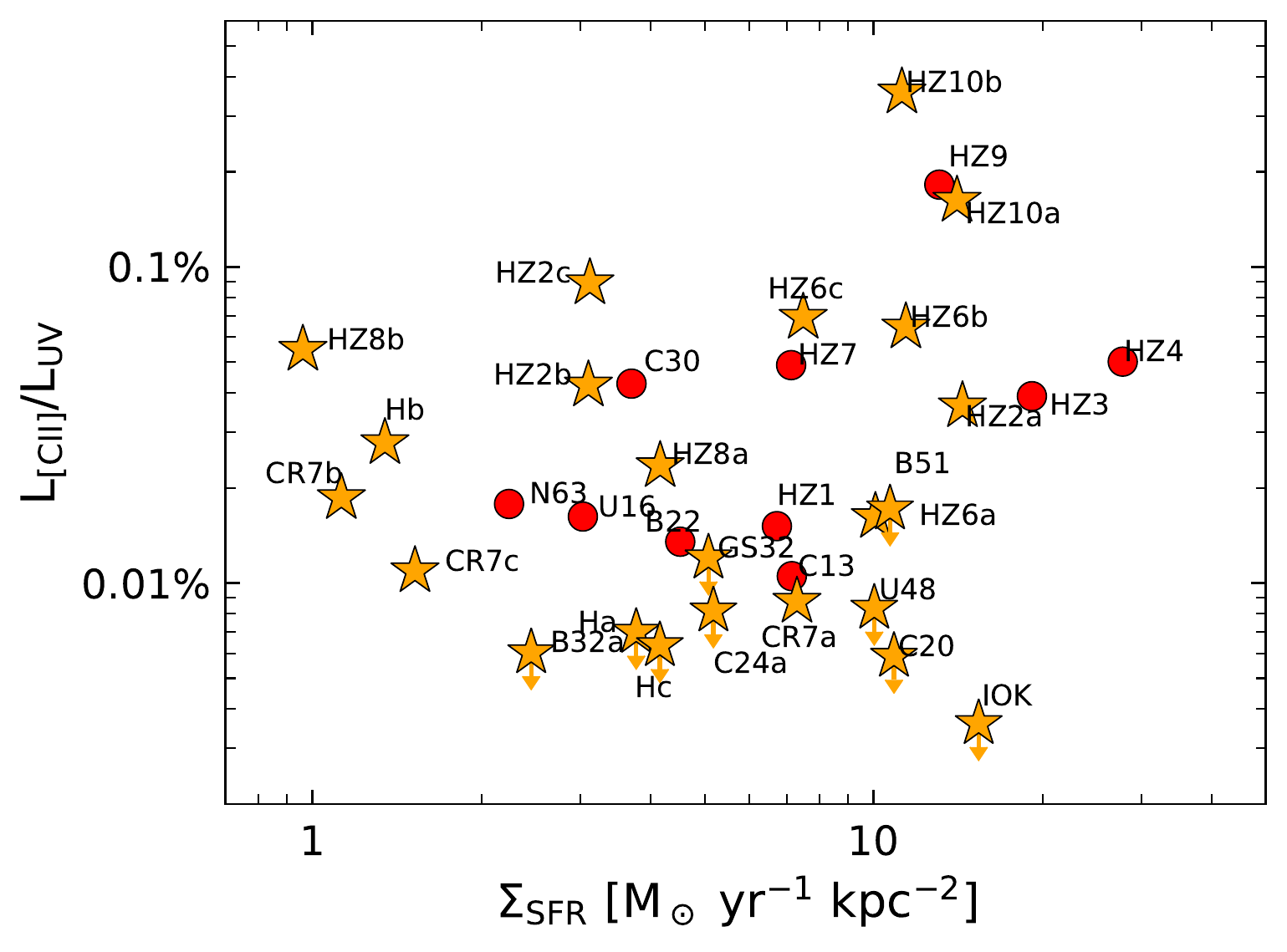}  %
\caption{\lcii/L$_{\rm UV}$ ratio as a function of the SFR surface density. Symbols are as in Figure~\ref{fig:fig2}.}
 \label{fig:fig5}
\end{figure}

However, the offset observed
in high-$z$ galaxies is significantly larger (one order of magnitude or more) than the one observed in local low metallicity galaxies, so other effects
are likely in place, which will be discussed in the next section.
Finally, still within the context of the surface brightness and SFR surface density, we mention that, for local galaxies,
\cite{Smith:2017} found a dependence between the \lcii/L$_{\rm FIR}$ ratio and the SFR surface density, suggesting that the \cii\ deficit increases strongly with increasing \esfr.  Therefore the \esfr\  dependence has a strong impact also on the  \mbox{\lcii-SFR} relation. 
In our sample, we cannot verify this dependence as most galaxies have only an upper limit on the L$_{\rm FIR}$. However, we can investigate  the relation  \lcii/L$_{\rm UV}$ and \esfr\, since the contribution of SFR$_{\rm FIR}$ to the total SFR is negligible (see discussion in Section~\ref{sec:SFRestimates}).
Figure~\ref{fig:fig5} shows the  observed \lcii/L$_{\rm UV}$ ratio spanning a range between 0.002\% to 0.4\% (over two orders of magnitude),
while the SFR surface density spans the range between 1 and 30 ${\rm M_{\odot} \ yr^{-1} \ kpc^{-2}}$.
There is only a very weak correlation between
\lcii/L$_{\rm UV}$ and \esfr, with very large dispersion (much larger than what observed locally for the
\lcii/L$_{\rm FIR}$ and \esfr relation). This indicates that in these high-$z$ systems the \lcii/L$_{\rm UV}$ line ratio (and \lcii/SFR) does not strongly depend on the
areal density with which galaxies form stars at $z>5$, and that other effects (metallicity, and other phenomena discussed in the next section) may contribute
to the very large dispersion.


\section{Discussion on the \cii-SFR scaling relations at z=5--7}\label{sec:discussion}

Clearly galaxies at $z>5$ behave differently, relative to their local counterparts, for what concerns the  \cii\ and SFR properties.
Summarising the finding of the previous sections:
1) both the \mbox{\lcii-SFR} relation and the \ecii--\esfr\ relation have a scatter much larger than the local relations;
2) contrary to some previous claims,  the \mbox{\lcii--SFR} relation is not offset relative to the local relation, while the \ecii--\esfr\ relation is clearly
offset by showing much lower \ecii \ relative to local relation;
3) the extension of the \cii\ emission is larger than the extent of the star formation traced by the UV emission.

The larger scatter observed in the SFR-\lcii\ relation is certainly
indicative of a broader range of properties spanned by such primeval galaxies relative to the local population.
Indeed, \mbox{high-$z$} cosmological simulations show  that the \cii\ emission strongly  depends on the gas metallicity, ionisation parameter, and  evolutionary stage of
the system and that all of these properties are expected to span a much broader range at high-$z$ with respect to local galaxies \citep{Vallini:2015,Vallini:2016,Vallini:2017, Pallottini:2015, Pallottini:2017, Olsen:2017, Katz:2017}.
Recently, \cite{Lagache:2017} investigated the expected dispersion of \lcii-SFR relation in the distant Universe by  using a semi-analytical model of galaxy formation
for a large sample of simulated galaxies at $z>4$. They found a \lcii-SFR  correlation with a large scatter of 0.4-0.8 dex, which is in agreement with our result. They claim that such large dispersion is associated to the combined effects of  different gas contents,  metallicities,  and interstellar radiation fields in the simulated high-$z$ galaxies. 

As we have shown, the mild anti-correlation with EW(Ly$\alpha$), as well as the analysis of some individual galaxies for which the metallicity has been estimated,
does suggest that the metallicity may play a role on the \cii\ emission (either simply in terms of carbon abundance and/or in terms of heating of the ISM through
photoelectric effect on dust grains, whose abundance scales with the metallicity). In particular, the lower metallicity of galaxies at $z>5$ 
can explain some of the scatter towards low \cii\ emission in both the \lcii-SFR relation and in the \ecii-\esfr\ relation. However, metallicity effects are
unlikely to explain the scatter towards high \cii\ emission in the  \lcii-SFR relation. Moreover, for what concerns the \ecii-\esfr\ relation, the offset
and large spread toward low \ecii\ is probably too large to be entirely ascribed to metallicity.

Large variations in ionisation parameter can also contribute to the spread in \cii\ emission \citep{Gracia-Carpio:2011, Katz:2017}. In particular, if \cii\ in
primeval galaxies  also traces circumgalactic gas in accretion and/or expelled from the galaxy, and excited by the UV radiation of the central galaxy,
this would result into the observed larger \cii\ sizes, hence lower \ecii \ and lower ionisation parameter, and the latter would increase the total \lcii \ relative to the local
relation. On the other hand, young compact star forming, primeval galaxies would be characterised by higher ionisation parameter, which would reduce the \cii\ emission,
hence contributing to the spread towards low \lcii.

As mentioned in the previous sections, yet another possibility is that the UV emission associated with some of the \cii\ clumps is heavily obscured. This would
explain the scatter above the \lcii-SFR relation, in the sense that for some of these systems the underlying SFR, as traced by the UV, is heavily underestimated. 
This can be the case for a few galaxies. However, as we mentioned, most of these systems show no or weak continuum dust emission, indicative of low dust content.

To make further progress additional data at other wavelength will be, in the future, extremely valuable. In particular, JWST will enable to identify obscured stellar components
as well as H$\alpha$ emission associated with star formation. ALMA observations of other transitions, such has [OIII]88$\mu$m (though observable only in some redshift ranges), has
proved extremely useful to constrain these scenarios \citep[e.g. ][]{Inoue:2016,Carniani:2017a}.

\section{Conclusions}\label{sec:conclusions}

In this work we have investigated the nature of the \cii\ emission in  star-forming galaxies at $z=5-7$. In particular, we have explored the positional offsets between  UV  and  FIR line emissions, the
presence of multiple components,
and the implications on the \cii-SFR scaling relations in the distant Universe, once the correct association between \cii\ and UV emission is properly taken into account.
We have performed our investigation in a sample of 29 $z>5$ ``normal'' star-forming galaxies (SFR $<$ 100 M$_{\odot}$~yr$^{-1}$) observed with ALMA in the \cii\ line. In addition to the re-analysis of archival objects,
we have also included  new ALMA observations targeting five star-forming galaxies at $z\sim6$ with SFR $\sim$ 10 \sfr, resulting into a new detection. Our mains results are:

\begin{itemize}


\item The continuum emission around 158 $\mu$m is not detected in most of the $z>5$ galaxies observed with ALMA, indicating a low dust content. By modelling the dust emission with a greybody spectrum
with dust temperature T$_{\rm d}$ = 30 K and emissivity index $\beta$ = 1.5, we have found that the SFR based on the FIR emission is, on average, lower than
the SFR measured from the  UV emission by at least a factor 0.6, but probably much more (due to several upper limits).

\item By accurately registering ALMA and NIR images, and by  kinematically discriminating multiple \cii\ components, our analysis has revealed that the \cii\ emission breaks into multiple
sub-components in 9 out of the 21  galaxies having \cii\ detections. In these nine targets we have observed the presence of 19 FIR-line emitting clumps. Only very few of these, if any, are associated
with the primary (brightest) UV counterpart, while the bulk of the \cii\ is associated with fainter UV components.
In only three cases (COSMOS24108, Himiko, and BDF3299) the shallow NIR images have not enabled us to detect the UV counterparts associated with some of the \cii\ clumps. 

\item  We  have studied the relation between \cii\  and SFR on the high-$z$ sample by  taking into account the presence of these sub-components and the proper associations
between \cii\ and UV components. The distribution of $z>5$ galaxies on the \lcii-SFR diagram follows the local relation, but the dispersion is 1.8 times larger than that observed in nearby galaxies.

\item The deviation from the local \lcii-SFR relation shows a weak anti-correlation with EW(Ly$\alpha$) though shallower and with larger dispersion than what found in other
studies that did not account for the multi-component nature of these systems.

\item  Most of the objects in the high-$z$ sample are spatially resolved in \cii\ and UV emission. 
The extension of the \cii\ emission is generally much larger than the extension of star forming regions traced by the UV emission.

\item In the \ecii-\esfr\ diagram $z>5$ galaxies are characterised by a large scatter with respect to local galaxies, and are mostly distributed below the local relation (i.e. fainter
\ecii\ at a given \esfr). 

\end{itemize}

We have suggested that a combination of different effects may be responsible for the different properties of high-$z$ galaxies in terms of \cii--SFR properties
relative to local galaxies. More specifically: 1) the low metallicity of high-$z$ galaxies may be responsible (also indirectly through the lower dust photoelectric heating) for part
of the scatter towards lower \cii\ emission relative to the local relations; 2) the presence of circumnuclear gas in accretion and/or expelled from the galaxy may be responsible
for the larger size in \cii\ relative to the SFR distribution and may also be responsible for the scatter of the \lcii-SFR distribution above the local relation
as a consequence of lower ionisation parameter; 3) in compact young star forming regions the increased ionisation parameter and higher gas density may be responsible for
the suppression of \cii\ for galaxies which are below the local relation; 4) dust obscuration may be responsible for both the different morphology between \cii\ and UV emission
and also for the scatter of sources above the local \lcii-SFR relation.

\section*{Acknowledgments}

This paper makes use of the following ALMA data: {\small ADS/JAO.ALMA\#2012.1.00719.S, ADS/JAO.ALMA\#2012.A.00040.S, ADS/JAO.ALMA\#2013.A.00433.S ADS/JAO.ALMA\#2011.0.00115.S, ADS/JAO.ALMA\#2012.1.00033.S, ADS/JAO.ALMA\#2012.1.00523.S, ADS/JAO.ALMA\#2013.1.00815.S, ADS/JAO.ALMA\#2015.1.00834.S., ADS/JAO.ALMA\#2015.1.01105.S, and ADS/JAO.ALMA\#2016.1.01240.S}
 which can be retrieved from the ALMA data archive: https://almascience.eso.org/ alma-data/archive. ALMA is a partnership of ESO (representing its member states), NSF (USA) and NINS (Japan), together with NRC (Canada) and NSC and ASIAA (Taiwan), in cooperation with the Republic of Chile. The Joint ALMA Observatory is operated by ESO, AUI/NRAO and NAOJ.
We are grateful to G. Jones to for providing the \cii\ flux maps of WHM5.
R.M. and S.C. acknowledge support by the Science and Technology Facilities Council (STFC). R.M. acknowledges ERC Advanced Grant 695671 ``QUENCH''. AF acknowledges support from the ERC Advanced Grant INTERSTELLAR H2020/740120.

\setlength{\labelwidth}{0pt}
\bibliographystyle{mn2e}
\bibliography{bibliography_multiclumps_morphology_in_highz_sf_galaxies.bib} 




\begin{center}

\begin{table*}
\caption{Star-forming galaxies at $z>5$}           
\label{tab:table1}  
\begin{threeparttable}    
\centering      
\begin{tabular}{l c c c c c c c c c c}    
\\
\hline
NAME  & ID & Redshift & log(L$_{\rm UV})$ & SFR$_{\rm UV}$ & log(L$_{\rm [CII]})$ & r$_{\rm UV}$ & r$_{\rm [CII]}$ & S$_{\rm mm}$ & ${\rm \log(L_{\rm FIR})}$ & \\
  &     &          &  [L$_\odot]$             & [\sfr] & [L$_\odot$]  & [kpc] & [kpc] & [$\mu$Jy] & [L$_\odot$] & \\
(a)  & (b) & (c) & (d) & (e) & (f) & (g) & (h) & (i) & (j) & (k)\\
\hline
& & & & & & & & &  \\
HZ8$^{*}$ & HZ8 & 5.1533 & $ 11.17 $ & $ 25.4 $ & $ 8.7 $ & $ - $ & $ - $ & $ <90 $ & $ <10.7 $ & [1,2,3]\\
\quad HZ8a & HZ8a & 5.1533 & $ 11.04 $ & $ 18.8 $ & $ 8.4 $ & $ 1.2 $ & $ 3.2 $ & $ <90 $ & $ <10.7 $ & [1,2,3]\\
\quad HZ8b & HZ8b & 5.1533 & $ 10.57 $ & $ 6.4 $ & $ 8.3 $ & $ 1.5 $ & $ 1.2 $ & $ <90 $ & $ <10.7 $ & [1,2,3]\\
HZ7 & HZ7 & 5.2532 & $ 11.05 $ & $ 19.3 $ & $ 8.7 $ & $ 1.0 $ & $ 1.9 $ & $ <108 $ & $ <10.8 $ & [1,2,3]\\
HZ6$^{*}$ & HZ6 & 5.2928 & $ 11.47 $ & $ 50.7 $ & $ 9.2 $ & $ - $ & $ - $ & $ 220 $ & $ 11.1 $ & [1,2,3]\\
\quad HZ6a & HZ6a & 5.2928 & $ 11.11 $ & $ 22.1 $ & $ 8.3 $ & $ 0.9 $ & $ <1.9 $ & $ 30 $ & $ 10.3 $ & [1,2,3]\\
\quad HZ6b & HZ6b & 5.2928 & $ 11.0 $ & $ 17.2 $ & $ 8.8 $ & $ 0.8 $ & $ 1.6 $ & $ 120 $ & $ 10.9 $ & [1,2,3]\\
\quad HZ6c & HZ6c & 5.2928 & $ 10.81 $ & $ 11.1 $ & $ 8.7 $ & $ 0.8 $ & $ 1.7 $ & $ 100 $ & $ 10.8 $ & [1,2,3]\\
HZ9 & HZ9 & 5.541 & $ 10.95 $ & $ 15.3 $ & $ 9.2 $ & $ 0.9 $ & $ 1.5 $ & $ 516 $ & $ 11.5 $ & [1,2,3]\\
HZ3 & HZ3 & 5.5416 & $ 11.08 $ & $ 20.6 $ & $ 8.7 $ & $ 0.7 $ & $ <2.3 $ & $ <153 $ & $ <11.0 $ & [1,2,3]\\
HZ4 & HZ4 & 5.544 & $ 11.28 $ & $ 32.7 $ & $ 9.0 $ & $ 0.7 $ & $ 1.4 $ & $ 202 $ & $ 11.1 $ & [1,2,3]\\
HZ10$^{*}$ & HZ10 & 5.6566 & $ 11.19 $ & $ 26.6 $ & $ 9.4 $ & $ - $ & $ - $ & $ 1261 $ & $ 11.9 $ & [1,2,3]\\
\quad HZ10a & HZ10a & 5.6566 & $ 11.14 $ & $ 23.7 $ & $ 9.3 $ & $ 1.1 $ & $ 1.6 $ & $ 630 $ & $ 11.6 $ & [1,2,3]\\
\quad HZ10b & HZ10b & 5.6566 & $ 10.23 $ & $ 2.9 $ & $ 8.8 $ & $ 0.7 $ & $ 1.5 $ & $ 630 $ & $ 11.6 $ & [1,2,3]\\
HZ2$^{*}$ & HZ2 & 5.6597 & $ 11.15 $ & $ 24.3 $ & $ 9.0 $ & $ - $ & $ - $ & $ <87 $ & $ <10.8 $ & [1,2,3]\\
\quad HZ2a & HZ2a & 5.6597 & $ 11.08 $ & $ 20.6 $ & $ 8.6 $ & $ 0.7 $ & $ 1.5 $ & $ <87 $ & $ <10.8 $ & This Work\\
\quad HZ2b & HZ2b & 5.6597 & $ 10.85 $ & $ 12.0 $ & $ 8.5 $ & $ 1.1 $ & $ <2.7 $ & $ <87 $ & $ <10.8 $ & This Work\\
\quad HZ2c & HZ2c & 5.6597 & $ 10.52 $ & $ 5.7 $ & $ 8.5 $ & $ 0.8 $ & $ <2.7 $ & $ <87 $ & $ <10.8 $ & This Work\\
HZ1 & HZ1 & 5.6885 & $ 11.21 $ & $ 28.5 $ & $ 8.4 $ & $ 1.2 $ & $ 0.9 $ & $ <90 $ & $ <10.8 $ & [1,2,3]\\
WMH5$^{*}$  & WMH5  & 6.0695 & $ 11.36 $ & $ 59.0 $ & $ 8.7 $ & $ - $ & $ - $ & $ 91 $ & $ 10.8 $ & [6,7]\\
\quad WMH5a & WMH5a & 6.0695 & $ <10.5 $ & $ <5.0 $ & $ 8.5 $ & $ <2.3 $ & $ 0.6 $ & $ 42 $ & $ 10.5 $ & [6,7]\\
\quad WMH5b & WMH5b & 6.0695 & $ 11.36 $ & $ 59.0 $ & $ 8.4 $ & $ <2.3 $ & $ 1.4 $ & $ 49 $ & $ 10.6 $ & [6,7]\\
NTTDF2313 & N23 & 6.07 & $ 10.85 $ & $ 12.0 $ & $ <7.7 $ & $ <1.4 $ & $ - $ & $ <54 $ & $ <10.6 $ & This Work\\
BDF2203 & B22 & 6.12 & $ 10.97 $ & $ 16.0 $ & $ 8.1 $ & $ 1.1 $ & $ 2.7 $ & $ <69 $ & $ <10.7 $ & This Work\\
CLM1 & CLM1 & 6.1657 & $ 11.37 $ & $ 60.0 $ & $ 8.4 $ & $ <2.3 $ & $ <1.8 $ & $ <78 $ & $ <10.8 $ & [6]\\
GOODS3203 & GS32 & 6.27 & $ 11.02 $ & $ 18.0 $ & $ <8.1 $ & $ 1.1 $ & $ - $ & $ <123 $ & $ <11.0 $ & This Work\\
COSMOS20521 & C20 & 6.36 & $ 10.89 $ & $ 14.0 $ & $ <7.7 $ & $ 0.7 $ & $ - $ & $ <60 $ & $ <10.7 $ & This Work\\
UDS4812 & U48 & 6.561 & $ 10.87 $ & $ 13.0 $ & $ <7.8 $ & $ 0.7 $ & $ - $ & $ <72 $ & $ <10.8 $ & This Work\\
Himiko$^{*}$ & H & 6.595 & $ 11.07 $ & $ 20.4 $ & $ 8.1 $ & $ - $ & $ - $ & $ <27 $ & $ <10.4 $ & [13,14]\\
\quad Himiko-a & Ha & 6.595 & $ 10.45 $ & $ 4.9 $ & $ <7.3 $ & $ 0.7 $ & $ - $ & $ <27 $ & $ <10.4 $ & [13,14]\\
\quad Himiko-b & Hb & 6.595 & $ 10.26 $ & $ 3.1 $ & $ 7.7 $ & $ 0.9 $ & $ <2.0 $ & $ <27 $ & $ <10.4 $ & [13,14]\\
\quad Himiko-c & Hc & 6.595 & $ 10.49 $ & $ 5.4 $ & $ <7.3 $ & $ 0.7 $ & $ - $ & $ <27 $ & $ <10.4 $ & [13,14]\\
\quad Himiko-Lya & HL & 6.595 & $ <10.31 $ & $ <3.5 $ & $ 7.9 $ & $ - $ & $ 3.4 $ & $ <27 $ & $ <10.4 $ & [13,14]\\
CR7$^{*}$ & CR7 & 6.604 & $ 11.2 $ & $ 27.1 $ & $ 8.3 $ & $ - $ & $ - $ & $ <21 $ & $ <10.2 $ & [10]\\
\quad CR7a & CR7a & 6.604 & $ 10.9 $ & $ 15.6 $ & $ 7.9 $ & $ 0.9 $ & $ 3.0 $ & $ <21 $ & $ <10.2 $ & [10]\\
\quad CR7b & CR7b & 6.604 & $ 10.2 $ & $ 2.9 $ & $ 7.5 $ & $ 0.9 $ & $ <2.2 $ & $ <21 $ & $ <10.2 $ & [10]\\
\quad CR7c & CR7c & 6.604 & $ 10.3 $ & $ 4.0 $ & $ 7.4 $ & $ 0.9 $ & $ 3.8 $ & $ <21 $ & $ <10.2 $ & [10]\\
COSMOS24108$^{*}$ & C24 & 6.6294 & $ 10.99 $ & $ 16.7 $ & $ 8.1 $ & $ - $ & $ - $ & $ <54 $ & $ <10.7 $ & [4]\\
\quad COSMOS24108a & C24a & 6.6294 & $ 10.99 $ & $ 16.7 $ & $ <7.9 $ & $ 1.1 $ & $ - $ & $ <54 $ & $ <10.7 $ & [4]\\
\quad COSMOS24108b & C24b & 6.6294 & $ <10.37 $ & $ <7.3 $ & $ 8.1 $ & $ - $ & $ 2.1 $ & $ <54 $ & $ <10.7 $ & [4]\\
UDS16291 & U16 & 6.6381 & $ 10.71 $ & $ 8.8 $ & $ 7.9 $ & $ 1.0 $ & $ 2.4 $ & $ <60 $ & $ <10.7 $ & [4]\\
NTTDF6345 & N63 & 6.701 & $ 10.95 $ & $ 15.3 $ & $ 8.2 $ & $ 1.5 $ & $ <1.7 $ & $ <48 $ & $ <10.6 $ & [4]\\
COS-2987030247  & C29 & 6.8076 & $ 11.11 $ & $ 23.0 $ & $ 8.6 $ & $ - $ & $ 3.1 $ & $ <75 $ & $ <10.8 $ & [5]\\
SDF46975 & S46 & 6.844 & $ 10.94 $ & $ 15.4 $ & $ <7.8 $ & $ <2.0 $ & $ - $ & $ <58 $ & $ <10.7 $ & [8]\\
COS-3018555981  & C30 & 6.854 & $ 11.04 $ & $ 18.8 $ & $ 8.7 $ & $ 1.3 $ & $ 2.6 $ & $ <87 $ & $ <10.9 $ & [5]\\
IOK-1 & IOK & 6.96 & $ 10.94 $ & $ 15.1 $ & $ <7.5 $ & $ 0.6 $ & $ - $ & $ <63 $ & $ <10.8 $ & [11]\\
BDF512 & B51 & 7.008 & $ 10.54 $ & $ 6.0 $ & $ <7.8 $ & $ 0.5 $ & $ - $ & $ <52 $ & $ <10.7 $ & [8]\\
BDF3299$^{*}$ & B32 & 7.107 & $ 0.0 $ & $ 6.4 $ & $ 7.8 $ & $ - $ & $ - $ & $ <23 $ & $ <10.3 $ & [8,9]\\
\quad BDF3299a & B32a & 7.107 & $ 10.52 $ & $ 5.7 $ & $ <7.3 $ & $ 0.9 $ & $ - $ & $ <23 $ & $ <10.3 $ & [8,9]\\
\quad BDF3299b & B32b & 7.107 & $ <10.0 $ & $ <1.7 $ & $ 7.8 $ & $ - $ & $ 1.0 $ & $ <23 $ & $ <10.3 $ & [8,9]\\
COSMOS13679 & C13 & 7.1453 & $ 10.91 $ & $ 13.9 $ & $ 7.9 $ & $ 0.8 $ & $ 1.4 $ & $ <42 $ & $ <10.6 $ & [4]\\
SXDF-NB1006-2 & SXDF & 7.212 & $ 11.06 $ & $ 8.7 $ & $ <7.9 $ & $ - $ & $ - $ & $ <42 $ & $ <10.6 $ & [12]\\\hline
\end{tabular}
\begin{tablenotes}\footnotesize

\item {\bf Notes}: 
{\bf (a)} Name of the source; the asterisk mark ($^{*}$) indicates that the source has a multi-clump morphology.
{\bf (b)} ID used to indicate the source in the figures of this paper.
{\bf (c)} Redshift of the galaxy (or system) inferred from Ly$\alpha$. 
{\bf(d)} Rest-frame UV luminosity at 1600\AA. 
{\bf(e)} SFR based on the UV emission: ${\rm log}$(SFR/\sfr)=${\rm log(L_{\rm UV}/erg \ s^{-1})-43.35}$ \citep{Murphy:2011, Hao:2011,Kennicutt:2012}
{\bf(f)} \cii\ luminosity. 
{\bf(g, h)} Half-light radius in kpc for UV and \cii\ emission. 
{\bf(i)} Continuum emission (or 3$\sigma$ upper limit) at rest-frame 158$\mu$m.
{\bf(j)} FIR luminosity (or 3$\sigma$ upper limit) estimated from ALMA observations.
{\bf(k)} References: [1] \cite{Capak:2015}, [2] \cite{Barisic:2017}, [3] \cite{Faisst:2017a},
[4] \cite{Willott:2015}, [5] \cite{Jones:2017}, [6] \cite{Ouchi:2013},
[7] \cite{Matthee:2017}, [8] \cite{Pentericci:2016}, [9] \cite{Smit:2017},
[10] \cite{Maiolino:2015}, [11] \cite{Ota:2014}, [12] \cite{Carniani:2017a},
[13] \cite{Inoue:2016}, [14] \cite{Carniani:2017}.

\end{tablenotes}
\end{threeparttable}
\end{table*}
\end{center}

\appendix

\label{appendix1}

\section{Additional ALMA data - observation and data reduction}\label{sec:appA}

\begin{table*}
 \centering
  \caption{UV and far-IR properties of the new five $z\sim6-7$ sources observed with ALMA. }\label{tab:appTab1}
  \begin{tabular}{lcccccccccc}
  \hline
   Name     & RA(J2000) & DEC(J200) &  z$_{\rm Ly\alpha}$$^a$        & SFR$_{\rm UV}^{b}$ & $\rm \nu _{obs}([CII])^c$ &
   beam$^d$ & t$_{\rm exp}$$^e$ & $\sigma _{\rm cont}^f$ &  $\sigma _{line}^g$ & \lcii$^h$ \\
     & [deg] & [deg] & & $\rm M_{\odot}~yr^{-1}$] & [GHz] & [min$''\times$maj$''$
	   ] & [hours] & [$\rm \mu$Jy] & [mJy] & [$\rm 10^7~L_{\odot}$]\\
 \hline
 NTTDF2313 & 181.3804 & -7.6935  & 6.07 & 12 & 268.817 & $\rm 0.98 \times 0.71$ & 0.7   & 18 & 0.15 & $<$4.5\\
 BDF2203 & 336.958 & -35.1472 & 6.12 &  16 & 266.93 & $\rm 1.90 \times 1.11$ &  0.4 & 23 & 0.2 & $12.5\pm2.5$ \\
 GOODS3203 & 53.0928 & -27.8826 & 6.27 & 18 & 250.236 & $1.26\times1.03$  & 0.1 & 41 & 0.4 & $<$12.0  \\
 COSMOS20521 & 150.1396 &  2.4269 & 6.36 & 14  & 258.225 & $\rm 1.46 \times 1.20$ & 
  0.8 & 20 & 0.15 &  $<$4.8\\
    UDS4821 &  34.4768 &  -5.24728 &   6.561 &   13  & 251.361 & $\rm 0.24 \times 0.22$ & 0.3 & 24 & 0.2 & $<$6.7 \\ 

 \hline
\end{tabular}
\\

Notes:
$^a$Redshift from either Ly$\alpha$ line or spectroscopic Lyman break. The uncertainty is $<0.04$.
$^b$ SFR inferred from the rest-frame UV continuum adopting the calibration discussed in \citet{Kennicutt:2012}.
$^c$Expected \cii\ frequency according to $z_{\rm Ly\alpha}$.  
$^d$ ALMA synthesised beam.
$^e$On-source integration time.
$^f$Sensitivity in  ALMA continuum map.
$^g$Sensitivity in  spectral channels of 100~km/s.
$^h$\cii\ luminosity. The upper limits on the \lcii\ are at 3$\sigma$, and are calculated on a width of 100~km/s. 

\label{tab_list_sources}
\end{table*}


The five $z\sim6$ star-forming galaxies listed in Table~\ref{tab_list_sources} were observed with ALMA in band 6 during Cycle 3 and Cycle 4 (program ID \#2015.1.01105.S and \#2015.1.01240.S).
ALMA observations were carried out with a semi-compact array configuration with angular resolutions ranging from 0.3\arcsec\ and 0.8\arcsec. The sources were observed for a total on source integration time of 0.1-0.8 h with a precipitable water vapour of 0.4-1.4 mm, depending on the specific observation. For each target we used four spectral windows (SPWs) set up in frequency division mode with a spectral resolution of $\sim$30 MHz ($\sim35$ km/s) and bandwidth of 1.875 GHz. One of the four SPWs was tuned to the expected frequency of the \cii\ line. The phase  of each observation was centred at the  NIR position of the respectively source.

J2248-3235, J0948-002, J1147-0724, J0239-0234, and J0552-3627 were observed as phase calibrator, respectively for the four sources. The flux calibrators were J2056-4714, Ganymede, J1229+0203, J1229+0623, and J0334-4008 while  bandpass calibrations were carried out through the observations of J2258-2758, J1058+0133, J1229+0203, J0238+1636, and J0522-3627.

ALMA observations were calibrated by using CASA software version v4.5.2 \citep{McMullin:2007}. Continuum and data cube images were obtained by using the CASA task {\sc clean} and natural weighting. The final  angular resolution and sensitivity reached in each set of data are listed in Table~\ref{tab:appTab1}.

We  registered NIR images to ALMA observations by matching the location of the foreground sources and ALMA calibrators to the position given by the GAIA Data Release 1 catalogue \citep{Gaia-Collaboration:2016}.

\begin{figure*}
\centering
\includegraphics[width=2\columnwidth]{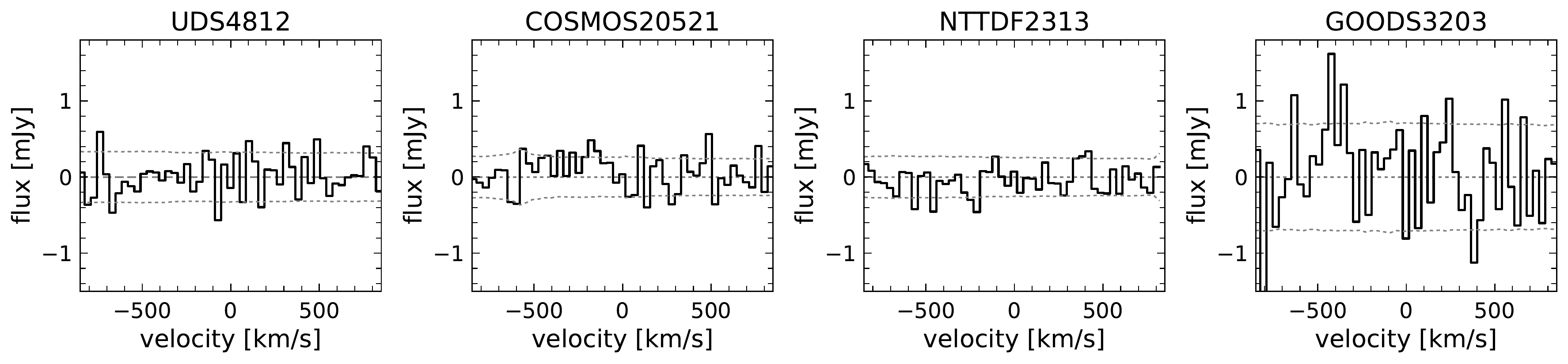}  %
\caption{ ALMA spectra of UDS4812, COSMOS20521, NTTDF2313, and GOODS3203. The spectra have been extracted at the location of the UV emission within a ALMA beam region. The velocity reference is set to the redshift inferred from spectroscopic rest-frame optical observations. The dotted grey lines shows the 1$\sigma$ and -1$\sigma$ }  
 \label{fig:appFig2}
\end{figure*}

As discussed in Section~\ref{sec:additional}, \cii\ emission was detected only in one target, BDF3203. 
Figures~\ref{fig:appFig1} and \ref{fig:appFig2} show the ALMA \cii\ spectra for the five star-forming galaxies extracted from a  region as large as the ALMA beam and centred at the location of the rest-frame UV regions.

\end{document}